# Dominance reversals: The resolution of genetic conflict and maintenance of genetic variation

Karl Grieshop[1,2,3*], Eddie K. H. Ho[4] and Katja R. Kasimatis[1,5]

[1]Department of Ecology and Evolutionary Biology, University of Toronto, Toronto, Canada

[2]Department of Molecular Biosciences, The Wenner-Gren Institute, Stockholm University, Stockholm, Sweden

[3]School of Biological Sciences, University of East Anglia, Norwich Research Park, Norwich, NR4 7TJ, United Kingdom

[4]Department of Biology, Reed College, Portland, Oregon, United States

[5]Department of Biology, University of Virginia, Charlottesville, Virginia, United States

ORCIDs:

https://orcid.org/0000-0001-8925-5066 (KG)

https://orcid.org/0000-0002-7692-394X (EKHH)

https://orcid.org/0000-0003-0025-5022 (KRK)

*Corresponding author:

Email: K.Grieshop@uea.ac.uk (KG)

**ABSTRACT**

Beneficial reversals of dominance reduce the costs of genetic trade-offs and can enable selection to maintain genetic variation for fitness. Beneficial dominance reversals are characterized by the beneficial allele for a given context (*e.g.* habitat, developmental stage, trait, or sex) being dominant in that context but recessive where deleterious. This context-dependence at least partially mitigates the fitness consequence of heterozygotes carrying one non-beneficial allele for their context and can result in balancing selection that maintains alternative alleles. Dominance reversals are theoretically plausible and are supported by mounting empirical evidence. Here we highlight the importance of beneficial dominance reversals as a mechanism for the mitigation of genetic conflict and review the theory and empirical evidence for them. We identify some areas in need of further research and development and outline three methods (dominance ordination, allele-specific expression, and allele-specific ATAC-Seq) that could facilitate the identification of antagonistic genetic variation. There is ample scope for the development of new empirical methods as well as reanalysis of existing data through the lens of dominance reversals. A greater focus on this topic will expand our understanding of the mechanisms that resolve genetic conflict and whether they maintain genetic variation.

# INTRODUCTION

Explaining the maintenance of genetic variation has been a mainstay pursuit in evolutionary biology since the modern synthesis [1–9]. One mechanism of particular interest is genetic trade-offs. A genetic trade-off occurs when alternative alleles are expressed in more than one context (*i.e.*, habitat, developmental stage, trait, or sex) with differing fitness effects between contexts. Alleles that are unambiguously beneficial across contexts can be fixed by positive directional selection. Those alleles that are unambiguously deleterious can be removed from the population. However, selection can maintain genetic variation when it favors alternative alleles in alternative contexts (e.g. antagonistic pleiotropy [10–13], sexually antagonistic selection [14,15], and some forms of spatially or temporally varying selection [16–19]) or can at least prolong the loss of additive genetic variance since antagonistic alleles will have longer transit times to fixation than unconditionally beneficial/deleterious alleles [15]. There is abundant evidence consistent with antagonistic genetic variation underlying some proportion of variance in fitness or its components due to trade-offs between alternative life history traits [20–25], tissue types [26,27], sexes [28–39], environments or generations [40–44].

Until and unless the lower net-fitness allele of these antagonistic polymorphisms is lost, this form of genetic variation will impose a genetic load [2,45–48] on population mean fitness, as selection is either maintaining or slowing the loss of alleles that harm the fitness of some individuals in some context(s). Theoretical and empirical evidence suggest there is strong selection pressure to resolve this form of genetic conflict [49–51].

One mechanism that partially resolves the genetic load imposed by antagonistic genetic variation is beneficial reversals of dominance, where alternative antagonistic alleles are dominant in their beneficial context and recessive in their deleterious context [10,14,17–19,52]. For example, a dominance reversed sexually antagonistic polymorphism would entail the female-beneficial (male-detrimental) allele having dominant fitness effects in heterozygous females but recessive fitness effects in heterozygous males, and the reverse for the male-beneficial (female-detrimental) allele. Similarly, under antagonistic pleiotropy between survival and reproduction, a heterozygous individual would have the harmful effects of each allele masked by fitness benefits of the other allele, one having dominant benefits to survival but recessive harmful effects on reproduction and vice versa for the other allele. Hence, context-dependent dominance reversals partially resolve the genetic conflict by allowing heterozygous individuals to be closer in fitness to that of the preferred homozygous genotype of each context, improving context-specific (and hence population) mean fitness – a reduction in the genetic load. Note that even complete dominance reversals can only ever partially resolve the genetic conflict, as there will always be both types of homozygotes in the 'wrong' contexts.

Whereas other forms of resolving genetic trade-offs such as gene duplication [53–56], sex-linkage [57–59], and epigenetics [60,61] may result in the loss of genetic variation, partial resolution by dominance reversal actually promotes the maintenance of the polymorphism [62]. Dominance reversal results in marginal overdominance (a net heterozygote advantage when averaged across contexts) or potentially overdominance (true heterozygote advantage) when the genetic trade-off is between individual fitness components [10], as in the Soay sheep ([24]; details in Empirical Examples). Higher (mean) heterozygote fitness stabilizes the polymorphism and can deterministically maintain this form of genetic variation for fitness ([10,14,17–19,52]). The strength of selection to resolve genetic trade-offs hinges on whether the cost of the conflict (magnitude of genetic load) is great enough and therefore the degree to which they have already been (partially) resolved by competing mechanisms [53]. For example, a dominance reversal that partially resolves a genetic trade-off would reduce the strength of selection for a gene duplication at that same polymorphic locus [53]. Hence, the timescales and constraints surrounding dominance reversals relative to other forms of resolution is likely highly consequential to the fate of genome architecture, genetic load, genetic variation, and the adaptive potential of the population.

Early skepticism over the plausibility of dominance reversals [12,63,64] may have delayed our understanding of this fundamental concept, but likely stemmed from the limited empirical evidence prior to the last decade [65,66]. For example, alternative variants of *Est*-4 in *Drosophila mojavensis* were shown to preferentially hydrolyze alternative esters and the heterozygotes faintly resemble the more efficient homozygote on each substrate [67] (table 1). Similarly, alternative *LDH* allozyme variants in the fish *Fundulus heteroclitus* showed opposite substrate affinity in alternative temperatures and heterozygote enzyme activity resembles the more efficient homozygote in each temperature [68] (table 1). And heterozygous amylase activity in *D*. *melanogaster* was closer to the low-activity homozygote in starch food and closer to the high-activity homozygote in normal food [69] (table 1). While these examples may seem only distantly related to organismal fitness in nature, recent empirical evidence is more convincing, putting dominance reversal back in the spotlight.

In this review, we describe the theory of dominance reversals, their ability to partially resolve genetic conflicts, and their relationship to genetic variation. We then highlight empirical examples consistent with dominance reversals across several sub-disciplines. Finally, we provide a forward look at identifying dominance reversals using quantitative genetics and gene expression analyses. Taken together, dominance reversals are not only plausible but supported by accumulating theory and empirical evidence, and their ability to partially resolve genetic trade-offs could render tests of dominance reversal (paired with validation tests) a powerful complement to existing methods of identifying antagonistic polymorphisms in the genome.

**Table 1. Empirical evidence for beneficial reversals of dominance.**

| Study | Species | Evidence | Context | Data | Method | Analysis |
|---|---|---|---|---|---|---|
| Zouros and Van Delden 1982[67] | *Drosophila mojavensis* | Single locus | Environments | Phenotypic | Crossing scheme | Enzyme assay |
| Place and Powers 1984[68] | *Fundulus heteroclitus* | Single locus | Environments | Phenotypic | Crossing scheme | Enzyme assay |
| Matsuo and Yamazaki 1986[69] | *Drosophila melanogaster* | Single locus | Environments | Phenotypic | Crossing scheme | Enzyme assay |
| Via *et al.* 2000[70] | *Acyrthosiphon pisum* | Polygenic | Environments | Fitness | Crossing scheme | Statistical modeling |
| Posavi *et al.* 2014[43] | *Eurytemora affinis* | Polygenic | Environments | Fitness component | Crossing scheme | Statistical modeling |
| Chen *et al.* 2015[71] | *Drosophila melanogaster* | Polygenic | Environments | Gene expression | Transcriptomic | Allele-specific expression |
| Johnston *et al.* 2013[24] | *Ovis aries* | Single locus | Traits | Fitness component | Pedigree / GWAS | Statistical modeling |
| Le Poul *et al.* 2014[72] | *Heliconius numata* | Inversion / Polygenic | Traits | Phenotypic | Crossing scheme | Image analysis |
| Gautier et al. 2018[73] | *Harmonia axyridis* | Single locus | Traits | Phenotypic | Crossing scheme | Image analysis |
| Mérot *et al.* 2020[74] | *Coelopa frigida* | Inversion | Traits / sexes | Fitness component | Experimental evolution | Statistical modeling / Numerical simulations |
| Jardine *et al.* 2021[75] | *Drosophila melanogaster* | Single locus | Traits | Fitness component | Genomics / phenotyping | Statistical modeling |
| Barson *et al.* 2015[35] | *Salmo salar* | Single locus | Sexes | Fitness component | Capture-recapture / GWAS | Statistical modeling |
| Grieshop and Arnqvist 2018[36] | *Callosobruchus maculatus* | Polygenic | Sexes | Fitness | Crossing scheme | Dominance ordination |
| Pearse *et al.* 2019[76] | *Oncorhynchus mykiss* | Inversion | Sexes | Fitness component | Capture-recapture | Statistical modeling |
| Geeta Arun *et al.* 2021[77] | *Drosophila melanogaster* | Polygenic | Sexes | Fitness component | Experimental evolution | Statistical modeling |
| Mishra *et al.* 2022[78] | *Drosophila melanogaster* | Polygenic | Sexes | Gene expression | Transcriptomic | Allele-specific expression |

## EVOLUTIONARY CAUSES OF DOMINANCE REVERSALS

The plausibility of dominance reversals starts with the plausibility of antagonistic polymorphisms. A direct extension of Fisher's geometric model [79–82] shows that for relatively well-adapted populations, the relatively rare instance of a mutation that would improve fitness in one context would tend to reduce fitness in a second context (assuming similar fitness landscapes between contexts and no mutational biases) [83–85]. Hence, antagonistic polymorphisms can readily arise but will tend to be unstable if simply additive.

Whether an antagonistic polymorphism is dominance reversed, or can become dominance reversed before it is lost, is clearly important and raises the question of how dominance reversals arise. The simplest answer is that antagonistic polymorphisms may be inherently dominance reversed – an inevitable outcome of overlapping and concave fitness functions (figure 1). Theory suggests that fitness functions are generally curved in the vicinity of their optima and that beneficial alleles should be dominant [83,86–90], with modest empirical support [91–96] (but see [97–100]). Hence, for a trait with context-antagonistic effects on fitness, a de facto dominance reversal will ensue at the underlying polymorphism(s) as long as the fitness functions overlap in their concave vicinities (figure 1D,iii; [17,19,88,101,102]). Still, the shape of the fitness landscape cannot be taken for granted. For example, two fitness functions could very plausibly overlap in their convex vicinities (e.g. overlapping tails of two Gaussian curves), sometimes resulting in net underdominance (heterozygote inferiority; figure S1E,v,vi), which would destabilize the polymorphism and rapidly fix one or the other alleles [47]. Whether the overlap between two real fitness functions occurs in their linear (figure 1C), concave (figure 1D), or convex (figure S1E) portions can only be resolved by empirical estimates of context-specific fitness functions. But dominance reversals need not rely on curved phenotype-fitness relationships since they can apparently be dominance reversed already at the phenotypic level (figure 1B), as in many of the empirical examples we discuss below (table 1). Note, however, that trait-level dominance reversals may or may not be dominance reversed for fitness, per se, depending on how the phenotype maps to fitness. In principle, dominance reversals at the phenotypic level could occur due to traits being threshold-like (non-linear genotype-phenotype relationship) [103], or could arise due to the explicit action of a dominance modifier [62].

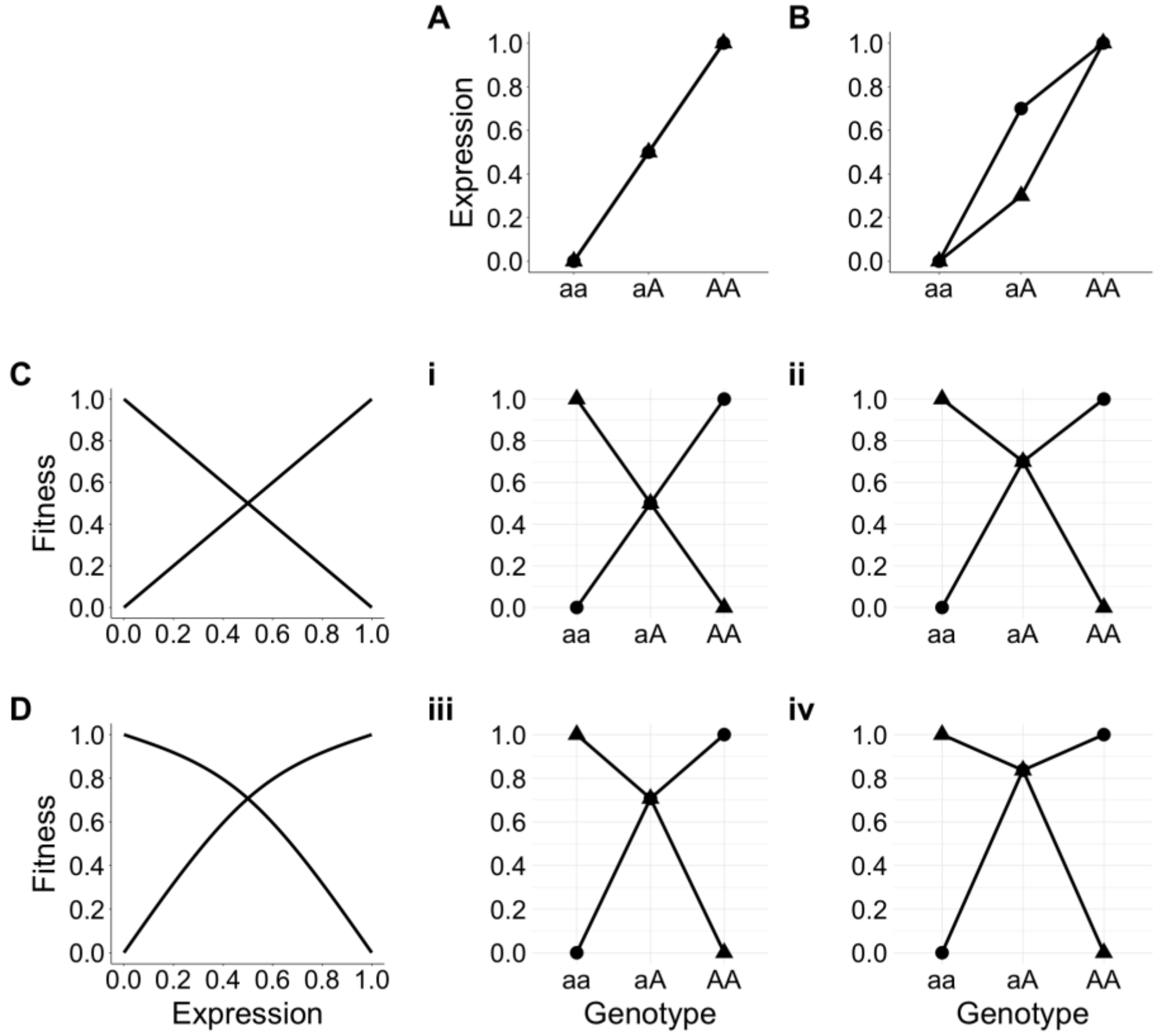

**Figure 1. Graphical representation of beneficial dominance reversals.** Circles versus triangles represent alternative sexes, environments, or generations (antagonistic pleiotropy scenarios not shown). We first map genotype to trait expression under two scenarios: **(A)** additivity for the phenotype, and **(B)** dominance reversal for the phenotype. Then we map trait expression to fitness under two scenarios of antagonistic selection: **(C)** additivity for fitness and **(D)** curved fitness functions overlapping in their concave vicinities. The ultimate effect of a genotype on fitness is given by the combination of the genotype-phenotype map (**A or B**) and the phenotype-fitness map (**C or D**) in the matrix of resultant genotype-fitness panels **i-iv**: **(i)** additivity, **(ii)** dominance reversal owing to the genotype-phenotype map (**ii**) or owing to the phenotype-fitness map (**iii**), and a greater magnitude of dominance reversal owing to their combined effects **(iv)**. See figure S1 for varying parameter settings, convex fitness functions, heterozygote inferiority, etc.

A dominance modifier could be any genetic or epigenetic process that can affect the dominance properties between alleles at an otherwise additive polymorphism. Unlike the de facto dominance reversals owing to the shape of the fitness landscape, a dominance modifier conferring a beneficial reversal of dominance would need to evolve in concert with the antagonistic polymorphism somehow. For example, a feature of pre-existent neutrally evolving regulatory network conducive to dominance reversed gene regulation could potentially facilitate an antagonistic mutation invading and sweeping to some intermediate equilibrium frequency, but there is currently no empirical or theoretical precedent for this. Instead, this has been approached from the starting assumption that an additive antagonistic polymorphism already exists with a sufficiently high frequency of heterozygotes. In that case, as Spencer and Priest [62] have demonstrated under sexually antagonistic selection, a dominance modifier mutation that partially resolves the genetic conflict of the additive antagonistic polymorphism is adaptive and will invade. This partial mitigation of the fitness costs paid by heterozygotes for carrying one of the deleterious alleles for their sex subsequently stabilizes the antagonistic polymorphism [62]. This theory is an extension of Otto and Bourguet's [104] theory, and the concept ultimately dates back more than one hundred years, even prior to (but including) the classical Fisher-Wright debate [62,105,106].

There are two outstanding issues with dominance reversals owing to a modifier. The first is that they either have to pre-date the antagonistic mutation (as described above) or rely on a sufficiently high proportion of heterozygotes at an additive antagonistic polymorphism (since the strength of selection for a modifier is roughly proportional to the probability of finding it in a heterozygote for the antagonistic site) [62,104]. But as discussed, most additive antagonistic polymorphisms will be progressing toward fixation for one or the other allele, so there may only be a narrow window of opportunity for a mutant dominance modifier to invade before drift or selection causes the antagonistic polymorphism to be lost. The second issue is that it is unclear what real-life gene regulatory phenomena might constitute a dominance modifier of this sort, which could bear consequence on its ability to evolve before the polymorphism is lost or whether there could be pre-existent neutrally evolving conditions conducive beneficial dominance reversal. Below, we provide a biophysically explicit proof-of-concept for what a dominance modifier could look like and analyze its evolution using forward-time individual-based population genetic simulations (Box 1; electronic supplementary material S2). Our example is a starting point for a theoretical approach that would be useful in assessing both of these outstanding issues, as well as whether antagonistic selection on a phenotype can maintain genetic variation [107], whether dominance reversals compete with gene duplications [108] other forms of resolution, and evaluating patterns in empirical data [78].

**Box 1. Dominance modifiers**

An allele of a transcription factor polymorphism can be dominant by having a lower mismatch to the downstream binding site or by having greater allele-specific concentration, either of which give a greater fractional occupancy of the binding site [109]. In the three-part linear regulatory network below, the sex-limited regulatory stimulus, $D$, and genes $A$ and $B$ are all unlinked, whereas the protein-coding domains of Genes $A$ and $B$, $A_i$ and $B_i$, are linked to their respective *cis*-regulatory binding sites, $\alpha_i$ and $\beta_i$. Only $\alpha_i$ and $\beta_i$ are allowed to mutate. The protein-coding domain of Gene $A$ harbors a sexually antagonistic polymorphism with male- and female-benefit alleles, $A_1$ and $A_2$, that respectively decrease and increase the expression level of Gene $B$, $[B]$. Those alleles are sexually antagonistic because the standardized $[B]$, $\varphi$, is under additive sexually antagonistic selection (as in figure 1C). (*We note there is precedent for gene expression levels showing opposite sign correlations with male versus female fitness in D. melanogaster* [110]).

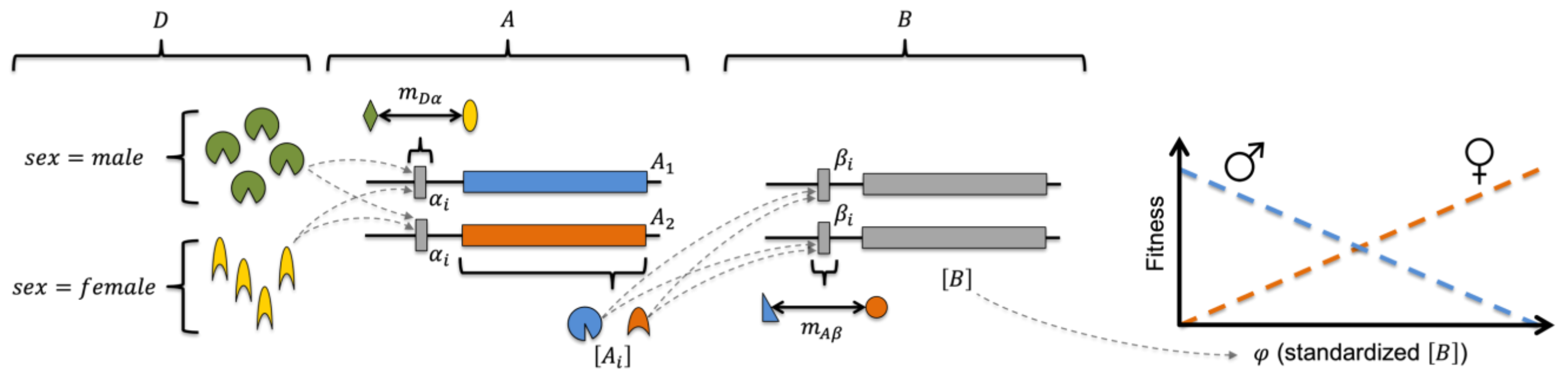

Arbitrarily, $A_2$ is intrinsically dominant over $A_1$ due to having a lower mismatch to the downstream substrate $\beta_i$ and therefore a greater fractional occupancy of that binding site, all else equal. But $A_1$ can be effectively dominant over $A_2$ if it can achieve a greater fractional occupancy via increased allele-specific concentration, $[A_i]$, determined as:

$$[A_i] \propto \left(\frac{[D]/k^{m_{D\alpha}(sex,\,\alpha_i)}}{1+2[D]/k^{m_{D\alpha}(sex,\,\alpha_i)}}\right), \tag{1}$$

where $[D]$ is the concentration of the sex-limited regulatory stimulus, $k$ is the stepwise change in the dissociation constant for $D$ binding to $\alpha_i$, and $m_{D\alpha}(sex, a_i)$ is the proportion of mismatched nucleotides between $D$ and $\alpha_i$ as a function of sex and $\alpha_i$. That is, in a heterozygous genotype $a_1A_1/a_2A_2$, mutations in $a_1$ that reduce $m_{D\alpha}(male, a_1)$ relative to $m_{D\alpha}(male, a_2)$ will increase the relative concentration of the linked male-benefit allele, $A_1$, in males.

Allowing these gene regulatory networks to evolve in forward-time individual-based population genetic simulations (electronic supplementary material S2) revealed that:

1. Selection favors the adaptive $\alpha_i A_i$ haplotypes that enable heterozygous males and females to plastically adjust the expression of Gene $B$ toward their sex-specific fitness optima (figure S2.4) because it partially resolves the genetic conflict (figure S2.5),
2. This dominance reversal in turn tends to maintain the focal sexually antagonistic polymorphism (figure S2.2),
3. The "dominance modifier" $\alpha_i$ cannot act alone and relies on the surrounding regulatory network, and
4. Our focal polymorphism exhibited a pattern of reversed allele-specific expression between the sexes (figure S2.6) – a detectable signature of dominance reversal [78].

## EVOLUTIONARY CONSEQUENCES OF DOMINANCE REVERSALS

Evidence of antagonistic genetic variation does not imply that the underlying polymorphisms would be maintained indefinitely under antagonistic balancing selection – most likely will not. While antagonistic mutations are inevitable (see previous section), to maintain an antagonistic polymorphism where the allelic effects on fitness are strictly additive (as in figure 1i) would require very strong and/or symmetric antagonistic selection between contexts [10,12,14,18,19,111,52]. However, even a partial dominance reversal would already cause a drastic expansion in the range of selection coefficients in each context that would deterministically maintain the polymorphism, including weak and/or asymmetric selection between contexts. This stabilizing effect of dominance reversal holds for genetic trade-offs between fitness components of individuals [10], between sexes [14,52] and between temporally and spatially varying environments [17–19,112] (especially in conjunction with other stabilizing mechanisms [113]).

The stability of a partially dominance reversed antagonistic polymorphisms owes to the same phenomenon that renders dominance reversal a resolution to conflict: the improved mean fitness of the heterozygotes. That is, above-intermediate mean heterozygote fitness not only improves population mean fitness (resolution of conflict), but it also increases the proportion of heterozygotes, which helps both alleles stay in circulation. When the alternative contexts of a genetic trade-off are not experienced by a single individual (*e.g.* sexes, environments, generations) this above-intermediate fitness of heterozygotes is a net heterozygote-advantage at the population level (marginal overdominance) [14,18,19], with the highest-fitness individuals still being the homozygotes in their preferred contexts (figure 1ii-iv). In cases where the alternative contexts are components of an individual's fitness (*i.e.* antagonistic pleiotropy) a dominance reversal can potentially result in true heterozygote advantage (overdominance) [10,24]. Indeed, evidence for overdominance may represent cases of dominance reversed antagonistic pleiotropy upon closer inspection, as in [24] (Table 1; details in Empirical Examples).

Whether overdominance or marginal overdominance, the consequence is an expansion the range of selection coefficients that would result in stable antagonistic polymorphism in the face of random genetic drift and/or imbalances in the strength of selection between contexts [10,14,17–19,52]. This expansion of the parameter space is especially impressive for small selection coefficients [52] such as those expected on average for each of the many loci presumed to underlie fitness and continuous traits [114]. As indirect empirical support for this notion, hundreds of SNPs with sexually antagonistic effects on fitness in *D. melanogaster* were significantly more likely to be trans-species polymorphisms shared with *D. simulans* relative to a randomly selected set of frequency-matched control sites [37], meaning that most of these sexually antagonistic polymorphisms arose prior to the speciation event between the two and were maintained over approximately one million years in both species ever since, which is highly unlikely to have occurred if they were simply additive.

## EMPIRICAL EXAMPLES

Surprisingly, few studies have explicitly set out to test for dominance reversals. We gathered empirical evidence consistent with dominance reversals spanning several sub-disciplines. In addition to the early evidence presented in the Introduction, below we highlight more recent examples of environment-specific, trait-specific, sex-specific, and multi-context dominance reversals. These examples emphasize the different methodologies, forms of evidence, and sub-disciplines that share this common interest. Many do not directly relate dominance reversal to the resolution of genetic conflict but rather to dominance reversal assisting the maintenance of antagonistic genetic variation; however, we note that the latter is owing to the former.

### Environment-specific dominance reversal

Posavi et al. [43] found evidence consistent with an environment-specific dominance reversal in the invasive copepod *Eurytemora affinis* (table 1). They derived two inbred strains from each of two high- and low-salinity environments and crossed them in a sex-specific full diallel cross [115] to compare survival of the within- versus between-environment $F_1$ offspring in each salinity environment. They found that while the within-environment crosses showed substantially lower survival in their ‘wrong’ salinity environments, the between-environment heterozygotes exhibited high survival in both salinities. Assuming high- and low-salinity adaptation in this system in largely governed by the same loci (though it may well not be), these results are consistent with dominance reversals maintaining genetic variation under antagonistic selection between environments. This finding is like the scenario depicted in figure 1B, which could facilitate the maintenance of genetic variation by spatially antagonistic selection. Via *et al.* [70] identify a similar example looking at the potential for ecological speciation between two populations of pea aphid (*Acyrthosiphon pisum*) that were alternatively adapted to alfalfa and clover (table 1). $F_1$ heterozygotes showed above-intermediate breeding values for fitness (per capita offspring/adult) on each host [70]. This dominance reversal is for an estimate of fitness and would likely generate marginal overdominance and balancing selection on the underlying polymorphisms [70], but the results are also consistent with heterosis (hybrid vigor) owing to divergent sets of unconditionally deleterious alleles between populations (see [102]).

Chen *et al*. [71] used allele-specific expression in two *D. melanogaster* strains and their $F_1$ hybrids reared in hot or cold temperatures to identify the dominance of parental alleles under different temperature environments. They identified 1,384 genes in which the opposite parental strain allele was dominant in the hot versus cold environment. However, because they assessed dominance by comparing $F_1$ hybrid effects to each homozygous parent, the inferred dominance effects could be due to background genetic differences between inbred parental strains. Still, some of these candidate genes likely do represent dominance reversals for expression and this set of genes exhibit patterns that are reminiscent of our dominance modifier model (Box 1). For example, we model a transcription factor polymorphism (Box 1) and their list of 1,384 genes was enriched for 13 different transcription factor binding sites [71]. Two of those 13 transcription factors were themselves putatively dominance reversed [71]. One transcription factor (*mip120*) exhibited *cis*-regulatory variation in the hot environment but *cis*- and *trans*-regulatory variation in the cold environment [71], which is consistent with our model of a dominance reversed transcription factor in that it requires the interplay between its *cis*-regulatory binding sites and a *trans*-acting regulatory stimulus (Box 1, electronic supplementary material S1). We note, however, that none of these putative dominance reversals have any known relationship to fitness or its components.

**Trait-specific dominance reversal**

Johnston *et al*.'s [24] study of the gene *RXFP2* in wild Soay sheep (*Ovis aries*) shows how true overdominance for fitness can emerge from dominance reversed fitness components under antagonistic pleiotropy. Homozygous $Ho^{+}Ho^{+}$ males have relatively high reproductive success and low survival, while $Ho^{P}Ho^{P}$ males have relatively low reproductive success and high survival. Heterozygous males have nearly equal reproductive success to $Ho^{+}Ho^{+}$ males and nearly equal survival to $Ho^{P}Ho^{P}$ males, which combine to yield overdominance for fitness [24] (table 1). This scenario of overdominance for fitness can occur when the genetic trade-off is between fitness components that combine to determine an individual's total fitness, even in the absence of overdominance for the fitness components ([10]). Jardine *et al*. [75] likewise identified a case of antagonistic pleiotropy between survival and reproduction likely acting to maintain short (*S*) and long (*L*) alleles of the classic *fruitless* (*fru*) gene in *D. melanogaster*. The 43 bp polymorphic indel was in a 1 kb region of the genome with elevated nucleotide diversity and Tajima's D (genomic signatures consistent with long-term balancing selection). Follow-up experiments showed that *S/S* flies had greater male mating success and lower larval survival relative to the *L* allele in its hemizygous arrangement (*i.e.* L/-, which we note is not the ideal comparison). Heterozygotes tended to have equally high male mating success to *S/S* flies, but also equally high larval survival to *L/-* flies (table 1). Hence the beneficial allele for each trait was dominant for that trait's expression in heterozygotes.

Color hierarchies can result in alternative alleles being dominant for alternative color patches, which can be thought of as different traits. For example, Le Poul *et al.* [72] investigated the individual genes lying within a super gene that underly a mimicry polymorphism for wing color pattern with important fitness consequences in the butterfly *Heliconius numata*. All eight within-population allele-pairs studied show opposite parental alleles being dominant for different color patches in $F_1$ offspring [72]. Perhaps the clearest case of a dominance reversal between color patches is between the *tar* and *arc* alleles, where the *tar* allele is dominant to the *arc* allele with respect to its large black patch but recessive with respect to its white patch [72]. These dominance reversals may assist the stable maintenance of these widespread and persistent mimicry polymorphisms by limiting the production of intermediate, non-mimetic individuals [72]. In a similar case, Gautier *et al.* [73] show that harlequin ladybird beetles (*Harmonia axyridis*) with alternative *cis*-regulatory alleles of the *pannier* gene have distinct color patterns on their elytra, and their heterozygous offspring exhibit a unique color pattern owing to alternative alleles being dominant in some color patches but recessive in others (table 1). This dominance reversal may contribute to the maintenance of alternative *pannier* alleles [73], as alternative color variants seem to be subject to seasonally fluctuating natural and sexual selection [116].

**Sex-specific dominance reversal**

One of the turning points in the growing appreciation for dominance reversals came from Barson *et al.*'s [35] study of the *VGLL3* gene in Atlantic salmon (*Salmo salar*), which explains 39% of phenotypic variation in age at sexual maturity. Alternative early- (*E*) and late-development (*L*) alleles are under sexually antagonistic selection with males preferring the former and females the latter, but heterozygous male and female development resembles that of the *EE* and *LL* homozygous genotype, respectively [35]. While this may depend on population- and age-specific genetic architecture [117,118], it nevertheless may still partially resolve the genetic conflict between the sexes over this this life history trait. In another example of a life history trait under sexually antagonistic selection, Geeta Arun *et al*. [77] revealed a polygenic signal of dominance reversal by challenging replicate populations of *D. melanogaster* to evolve in response to pathogenic infection. After 65-75 generations of experimental evolution, the progeny of crosses between populations both within and between treatments were challenged to survive infection. The between-treatment male and female progeny were not perfectly intermediate between the within-infected and within-control crosses, rather, female heterozygotes exhibited above-intermediate survivorship (closer to that of the infected group) while males showed below-intermediate survivorship [77]. The results were consistent across all four replicate pairs of the experimental evolution program, suggesting a trade-off between immunocompetence and some other fitness-related trait in males. Both studies represent examples of a sex-specific dominance reversal at the phenotypic level (as in Fig 1B), which would likely assist in the stable maintenance of the underlying genetic variation via sexually antagonistic selection.

Grieshop and Arnqvist [36] uncovered a polygenic signal of sex-specific dominance reversal for fitness in seed beetles (*Callosobruchus maculatus*). They used a full diallel cross among 16 strains and obtained replicated estimates of sex-specific competitive lifetime reproductive success for all combinations. They then used the “array covariances” [115,119] to order the strains in order of how dominant their fixed allelic variation was relative to one another, separately for the male and female data. They found a negative genetic correlation between the male and female dominance ordination among strains, implying that strains whose fixed genetic variation tended to be dominant to that of other strains in one sex was recessive to that of other strains when measured in the opposite sex, a polygenic signal of sex-specific dominance reversal for fitness [52]. Mishra et al.’s [78] recent study of allele-specific expression in *D. melanogaster* also identified evidence of polygenic sex-specific dominance reversal. In 176 of 3796 quality-controlled genes, opposite alleles were significantly more highly expressed in opposite sexes (with as many as 26/176 representing false positives) [78]. This sex-specific dominance reversal for expression is consistent with a pattern predicted by our model (Box 1, figure S1.6), but as with Chen et al.'s [71] study they have no known relationship to fitness or its components.

### Multi-context dominance reversal

The final two studies we highlight both identified dominance reversals in major-effect autosomal inversion polymorphisms that underlie life history traits with sex-specific fitness optima. Both examples involve the interplay between life history trade-offs, environmental effects, and sex- or trait-specific dominance reversals combining to consequently maintain genetic variation at these fitness-determining inversions. Pearse *et al*. [76] investigated a large supergene (*Omy05*) in rainbow trout (*Oncorhynchus mykiss*), where the ancestral and rearranged alleles represent a polymorphism that underlies an alternative migratory-based reproductive strategy subject to environmentally-dependent and sexually antagonistic selection [76,120,121]. They found that the statistical model of best fit to explain their capture-recapture data was one that allowed for sex-specific dominance, which has therefore likely assisted in the maintenance of this major-effect multivariate antagonistic polymorphism for approximately 1.5 million years [76]. Similarly, Mérot *et al*. [74] examined a large inversion in the seaweed fly *Coelopa frigida*. The allelic effects of this inversion represent an environmentally dependent survival/reproduction trade-off that results in overdominance for fitness (as with other examples above), with their experimental evolution data suggesting that the overdominance emerges not only due to varying strengths and directions of dominance effects between life history traits but also between the sexes [74]. These examples highlight how multiple forms of antagonistic selection and dominance reversal can act in concert to maintain genetic variation.

## DETECTING SIGNATURES OF DOMINANCE REVERSALS

It is a longstanding problem to confidently detect antagonistic balancing selection using traditional methods [113,122–127] driving various creative analyses of polygenic signals [37,38,110,128–131]. For example, Ruzicka *et al*.'s [37] GWAS would have overlooked many sexually antagonistic SNPs (none being significant after Bonferroni correction) if not for reducing their significance threshold and testing for trans-species polymorphism (see end of *Evolutionary Consequences* section, above). Their method was optimized for detecting additive signals [37], meaning that the effects would have needed to be strong enough and/or additive enough to be detected through the noise of the unmodelled dominance effects predicted of such sites. Interestingly, there is very little overlap between Ruzicka *et al.*'s [37] list of candidate sexually antagonistic genes and Mishra *et al.*'s [78] list of genes with dominance reversed allele-specific expression between the sexes, suggesting the methods may complement one another. Similarly, Grieshop and Arnqvist [36] saw only a modest signal of additive sexually antagonistic effects using traditional quantitative genetic variance partitioning, but a definitive and strong signature of sex-reversed dominance ordination among strains using the same data. We propose that tests for dominance reversal – the signature of partially resolved antagonistic polymorphisms – paired with follow-up fitness validation tests would offer a powerful complementary approach to methods that aim to detect additive signals of antagonistic polymorphisms.

Besides those of empirical examples above (Table 1), three methods of testing for dominance reversals seem particularly promising: dominance ordination, allele-specific expression, and allele-specific ATAC-Seq. While this list does not capture all potential approaches (e.g. [132,133]), we believe it represents some of the most promising newer methods. There are likely many existing data sets in the medical and agricultural literature that are ripe for reanalysis through the lens of dominance reversal using one or more of these methods.

Dominance ordination (explained in greater detail in electronic supplementary material S2) uses the "array covariances" obtained from quantitative genetic data of a full diallel cross [115,119] to ordinate a panel of inbred strains from those whose fixed allelic variation (averaged across the genome) is mostly dominant over that of the other strains to those whose fixed alleles are mostly recessive to that of the other strains [36]. This process can be done separately for alternative contexts. The null expectation for underlying loci is that alternative alleles are either unconditionally dominant or unconditionally recessive, which would cause the strains to be in approximately the same order along the dominant-recessive continuum in both contexts. This null expectation would be detectable as a positive genetic correlation between contexts among the strain-specific array covariances ([36]; figure S3.2C). By contrast, a negative cross-context genetic correlation among the array covariances would indicate that strains tend to be fixed for alleles that are dominant in one context but recessive in the other ([36]; figure S3.2D), meaning that the underlying polymorphisms are dominance reversed. This method is currently best utilized as a quantitative genetic test of dominance reversal and does not reveal the causal polymorphisms underlying the signal. Possible avenues for further development of this method include extending it to other breeding designs and pedigree data [115] and/or applying the dominance ordination step at the level of chromosomes, linkage-blocks, or SNPs rather than strains or families.

By contrast, allele-specific expression analyses can identify specific genes with dominance reversed expression between alternative life stages, traits, tissues, sexes, or environments. Sites with context-dependent allelic imbalance in expression represent candidate regions of the genome harboring antagonistic polymorphisms, where further work would be required to identify candidate SNPs or indels as well as to validate their effects on fitness or fitness traits. We note that many allele-specific expression studies utilize crosses between compatible species because it increases the number of mRNA reads that can be confidently distinguished as having been inherited from one parent or the other (see below), but such data would not be relevant to the concept of dominance reversals resolving genetic conflicts (or maintaining genetic variation), as many of the variable sites would represent fixed differences between species. There are broadly two alternative allele-specific expression designs to identify dominance reversals: the "$F_1$ hybrids" approach and the "common reference" approach [134]. Under both designs, transcriptomic reads from maternally and paternally inherited alleles can be identified and quantified by mapping $F_1$ heterozygote reads back to both parental genotype-specific reference genomes [134,78]. Careful quality control and data filtering must be applied to rule out confounding parental effects and other sources of mapping bias (see [78]).

Once completed, relative read coverage between alternative alleles reveals allele-specific expression, and reversed allelic imbalance between contexts represents dominance reversed gene expression (figure S2.6). Genes should only be considered as true positive reversals of allelic imbalance if the magnitude of allelic imbalance is significant in both contexts independently. A statistically significant allele-by-context interaction term could include genes that exhibit allelic imbalance in one context but not the other, which would indicate context-dependent dominance but not a dominance reversal, per se. This method could be enhanced by using long-read technologies such as PacBio, Oxford Nanopore or Iso-Seq [135–139] or linked-read sequencing applied to transcriptomic data [140] to better assign reads to the correct parental genome. In principle, single-cell technologies [137,141] could be integrated within the allele-specific expression methods to reveal the full atlas of genes that are dominance reversed across all tissue types.

ATAC-Seq (assay for transposase-accessible chromatin with sequencing) could potentially reveal dominance reversed non-coding regions of the genome. ATAC-Seq identifies the genome-wide regions of organized DNA (*i.e.* chromatin) that are "open" for transcription (*i.e.* euchromatin) by sequencing only the unwound euchromatin to build a database of active loci [142,143]. ATAC-Seq requires less tissue, time, troubleshooting, and cost than CHIP-seq, while providing more information on the active loci, promoters, and enhancers genome-wide. Since the euchromatin or heterochromatin state of DNA varies across contexts [144], allele-specific ATAC-Seq [145] could in principle reveal non-coding regions that are dominance reversed for transcriptional availability, including loci that may regulate dominance reversed allele-specific gene expression (such as the cis-regulatory binding sites in Box 1, S2). ATAC-Seq also reveals transcription factor binding site motifs [146] that could be integrated with the analysis of allele-specific expression data [145] to identify the regulatory networks that govern dominance reversed gene expression. These results could be assessed for the extent to which certain regulatory networks are predisposed to maintaining antagonistic polymorphisms via dominance reversal and could be evaluated across different classes of genetic trade-offs to assess the extent of overlap in regulatory sites and networks that are conducive to maintaining polymorphisms under different forms of antagonistic selection. Again, dominance reversed allele-specific expression and ATAC-Seq findings would have no functional or fitness basis on their own and should be interpreted with caution and/or validated by manipulative experimentation.

## CONCLUSIONS

Beneficial reversals of dominance partially resolve genetic conflicts by improving population mean fitness, with the consequence of maintaining genetic variation. Studying whether different mechanisms compete or coordinate with dominance reversal to resolve genetic conflict has important implications for the evolution of genetic architecture and maintenance of genetic variation. Antagonistic genetic variation is likely to be enriched for dominance reversals because non-dominance-reversed antagonistic polymorphisms are sensitive to being lost by drift or unequal strengths of selection between contexts. We suggest that dominance ordination, allele-specific expression, and allele-specific ATAC-Seq represent promising methods of testing for these signatures of dominance reversal that are predicted to accompany antagonistic polymorphisms. There is ample scope for developing methods to identify signatures of dominance reversal and there are likely also many suitable data sets that already exist. We hope this article reignites research interest in dominance reversal as it bears broad and pivotal consequences to several areas of evolutionary biology.

## ACKNOWLEDGMENTS

We thank J.A. Ågren, G. Arnqvist, and L. Rowe for the opportunity to contribute to this special issue. We thank two very helpful anonymous reviewers for their careful and constructive comments. We thank A.F. Agrawal, T. Connallon, and A.D. Cutter for comments on earlier drafts of the manuscript. This research was supported by the Swedish Research Council (2018-06775 to KG), the University of Toronto's Faculty of Arts and Science (Postdoctoral Fellowship to KG), and the National Sciences and Engineering Research Council (Banting Postdoctoral Fellowship to KRK).

## AUTHOR CONTRIBUTIONS

KG and KRK conceived of the article. KG did the literature search and wrote the original draft of the manuscript. KG and EKHH conducted the novel theory presented in Box 1 (S1), with EKHH solely responsible for the Python coding. All authors contributed to editing the manuscript.

## COMPETING INTERESTS

The authors declare no competing interests.

## DATA ARCHIVING

The underlying Python code (for Box 1, electronic supplementary material S1) is available at https://gitfront.io/r/user-5416399/Ym11RGUhTp9L/simDominanceModifier/.

## REFERENCES

1. Wright S. 1935 Evolution in populations in approximate equilibrium. *J. Genet.* **30**, 257–266. (doi:10.1007/BF02982240)

2. Dobzhansky T. 1955 A review of some fundamental concepts and problems of population genetics. *Cold Spring Harb. Symp. Quant. Biol.* **20**, 1–15. (doi:10.1101/sqb.1955.020.01.003)

3. Kimura M. 1958 On the Change of Population Fitness by Natural Selection. *Heredity* **12**, 23.

4. R. C. Lewontin. 1974 *The genetic basis of evolutionary change*. New York, NY: Columbia University Press.

5. Barton NH, Turelli M. 1989 Evolutionary Quantitative Genetics: How Little Do We Know? *Annu. Rev. Genet.* **23**, 337–70.

6. Charlesworth B, Hughes KA. 2000 The maintenance of genetic variation in life-history traits. In *Evolutionary Genetics* (eds RS Singh, CB Krimbas), pp. 369–392. Cambridge: Cambridge University Press.

7. Barton NH, Keightley PD. 2002 Understanding quantitative genetic variation. *Nat. Rev. Genet.* **3**, 11–21. (doi:10.1038/nrg700)

8. Mitchell-Olds T, Willis JH, Goldstein DB. 2007 Which evolutionary processes influence natural genetic variation for phenotypic traits? *Nat. Rev. Genet.* **8**, 845–856. (doi:10.1038/nrg2207)

9. Charlesworth B. 2015 Causes of natural variation in fitness: Evidence from studies of Drosophila populations. *Proc. Natl. Acad. Sci.* **112**, E1049–E1049. (doi:10.1073/pnas.1502053112)

10. Rose MR. 1982 Antagonistic pleiotropy, dominance, and genetic variation. *Heredity* **48**, 63–78. (doi:10.1038/hdy.1982.7)

11. Rose MR. 1985 Life history evolution with antagonistic pleiotropy and overlapping generations. *Theor. Popul. Biol.* **28**, 342–358. (doi:10.1016/0040-5809(85)90034-6)

12. Curtsinger JW, Service PM, Prout T. 1994 Antagonistic Pleiotropy, Reversal of Dominance, and Genetic Polymorphism. *Am. Nat.* **144**, 210–228.

13. Brown KE, Kelly JK. 2018 Antagonistic pleiotropy can maintain fitness variation in annual plants. *J. Evol. Biol.* **31**, 46–56. (doi:10.1111/jeb.13192)

14. Kidwell JF, Clegg MT, Stewart FM, Prout T. 1977 REGIONS OF STABLE EQUILIBRIA FOR MODELS OF DIFFERENTIAL SELECTION IN THE TWO SEXES UNDER RANDOM MATING. *Genetics* **85**, 171–183. (doi:10.1093/genetics/85.1.171)

15. Connallon T, Clark AG. 2012 A General Population Genetic Framework for Antagonistic Selection That Accounts for Demography and Recurrent Mutation. *Genetics* **190**, 1477–1489. (doi:10.1534/genetics.111.137117)

16. Haldane JBS, Jayakar SD. 1963 Polymorphism due to selection of varying direction. *J. Genet.* **58**, 237–242. (doi:10.1007/BF02986143)

17. Gillespie JH, Langley CH. 1974 A general model to account for enzyme variation in natural populations. *Genetics* **76**, 837–848.

18. Hedrick PW. 1976 Genetic variation in a heterogeneous environment. II. Temporal heterogeneity and directional selection. *Genetics* **84**, 145–157.

19. Gillespie JH. 1978 A general model to account for enzyme variation in natural populations. V. The SAS-CFF model. *Theor. Popul. Biol.* **14**, 1–45. (doi:10.1016/0040-5809(78)90002-3)

20. Simmons MJ, Preston CR, Engels WR. 1980 Pleiotropic effects on fitness of mutations affecting viability in Drosophila melanogaster. *Genetics* **94**, 467–475.

21. Rose MR, Charlesworth B. 1981 Genetics of life history in Drosophila melanogaster. I. Sib analysis of adult females. *Genetics* **97**, 173–186.

22. Reznick D. 1985 Costs of reproduction: an evaluation of the empirical evidence. *Oikos* , 257–267.

23. Scheiner SM, Caplan RL, Lyman RF. 1989 A search for trade-offs among life history traits in Drosophila melanogaster. *Evol. Ecol.* **3**, 51–63.

24. Johnston SE, Gratten J, Berenos C, Pilkington JG, Clutton-Brock TH, Pemberton JM, Slate J. 2013 Life history trade-offs at a single locus maintain sexually selected genetic variation. *Nature* **502**, 93–95. (doi:10.1038/nature12489)

25. Oakley CG, Ågren J, Atchison RA, Schemske DW. 2014 QTL mapping of freezing tolerance: links to fitness and adaptive trade-offs. *Mol. Ecol.* **23**, 4304–4315. (doi:10.1111/mec.12862)

26. Kotrschal A, Rogell B, Bundsen A, Svensson B, Zajitschek S, Brännström I, Immler S, Maklakov AA, Kolm N. 2013 Artificial Selection on Relative Brain Size in the Guppy Reveals Costs and Benefits of Evolving a Larger Brain. *Curr. Biol.* **23**, 168–171. (doi:10.1016/j.cub.2012.11.058)

27. Chen H, Jolly C, Bublys K, Marcu D, Immler S. 2020 Trade-off between somatic and germline repair in a vertebrate supports the expensive germ line hypothesis. *Proc. Natl. Acad. Sci.* **117**, 8973–8979. (doi:10.1073/pnas.1918205117)

28. Rice W. 1992 Sexually antagonistic genes: experimental evidence. *Science* **256**, 1436–1439. (doi:10.1126/science.1604317)

29. Chippindale AK, Gibson JR, Rice WR. 2001 Negative genetic correlation for adult fitness between sexes reveals ontogenetic conflict in Drosophila. *Proc. Natl. Acad. Sci. U. S. A.* **98**, 1671–1675. (doi:10.1073/pnas.041378098)

30. Fedorka KM, Mousseau TA. 2004 Female mating bias results in conflicting sex-specific offspring fitness. *Nature* **429**, 65–67. (doi:10.1038/nature02492)

31. Foerster K, Coulson T, Sheldon BC, Pemberton JM, Clutton-Brock TH, Kruuk LEB. 2007 Sexually antagonistic genetic variation for fitness in red deer. *Nature* **447**, 1107–1110. (doi:10.1038/nature05912)

32. Delcourt M, Blows MW, Rundle HD. 2009 Sexually antagonistic genetic variance for fitness in an ancestral and a novel environment. *Proc. R. Soc. B Biol. Sci.* **276**, 2009–2014. (doi:10.1098/rspb.2008.1459)

33. Delph LF, Andicoechea J, Steven JC, Herlihy CR, Scarpino SV, Bell DL. 2011 Environment-dependent intralocus sexual conflict in a dioecious plant. *New Phytol.* **192**, 542–552. (doi:10.1111/j.1469-8137.2011.03811.x)

34. Berger D, Grieshop K, Lind MI, Goenaga J, Maklakov AA, Arnqvist G. 2014 Intralocus Sexual Conflict and Environmental Stress. *Evolution* **68**, 2184–2196. (doi:10.1111/evo.12439)

35. Barson NJ *et al.* 2015 Sex-dependent dominance at a single locus maintains variation in age at maturity in salmon. *Nature* **528**, 405–408. (doi:10.1038/nature16062)

36. Grieshop K, Arnqvist G. 2018 Sex-specific dominance reversal of genetic variation for fitness. *PLOS Biol.* **16**, e2006810. (doi:10.1371/journal.pbio.2006810)

37. Ruzicka F, Hill MS, Pennell TM, Flis I, Ingleby FC, Mott R, Fowler K, Morrow EH, Reuter M. 2019 Genome-wide sexually antagonistic variants reveal long-standing constraints on sexual dimorphism in fruit flies. *PLOS Biol.* **17**, e3000244. (doi:10.1371/journal.pbio.3000244)

38. Ruzicka F, Holman L, Connallon T. 2022 Polygenic signals of sex differences in selection in humans from the UK Biobank. *Plos Biol.* **20**, e3001768.

39. Rusuwa BB, Chung H, Allen SL, Frentiu FD, Chenoweth SF. 2022 Natural variation at a single gene generates sexual antagonism across fitness components in Drosophila. *Curr. Biol.* **32**, 3161-3169. e7.

40. Powell JR. 1971 Genetic polymorphisms in varied environments. *Science* **174**, 1035–1036.

41. Levinton J. 1973 Genetic variation in a gradient of environmental variability: marine bivalvia (Mollusca). *Science* **180**, 75–76.

42. Mojica JP, Lee YW, Willis JH, Kelly JK. 2012 Spatially and temporally varying selection on intrapopulation quantitative trait loci for a life history trade-off in *Mimulus guttatus*. *Mol. Ecol.* **21**, 3718–3728. (doi:10.1111/j.1365-294X.2012.05662.x)

43. Posavi M, Gelembiuk GW, Larget B, Lee CE. 2014 Testing for beneficial reversal of dominance during salinity shifts in the invasive copepod *Eurytemora affinis* , and implications for the maintenance of genetic variation: TESTING FOR REVERSAL OF DOMINANCE. *Evolution* **68**, 3166–3183. (doi:10.1111/evo.12502)

44. Troth A, Puzey JR, Kim RS, Willis JH, Kelly JK. 2018 Selective trade-offs maintain alleles underpinning complex trait variation in plants. *Science* **361**, 475–478.

45. Crow JF. 1958 SOME POSSIBILITIES FOR MEASURING SELECTION INTENSITIES IN MAN. *Hum. Biol.* **30**, 1–13.

46. Dobzhansky T. 1957 Genetic loads in natural populations. *Science* **126**, 191–194.

47. Charlesworth B, Charlesworth D. 2010 *Elements of evolutionary genetics*. Greenwood Village, CO: Roberts and Company Publishers.

48. Whitlock M, Davis B. 2011 Genetic Load. In *eLS* (ed John Wiley & Sons, Ltd), p. a0001787.pub2. Chichester, UK: John Wiley & Sons, Ltd. (doi:10.1002/9780470015902.a0001787.pub2)

49. Bonduriansky R, Chenoweth SF. 2009 Intralocus sexual conflict. *Trends Ecol. Evol.* **24**, 280–288. (doi:10.1016/j.tree.2008.12.005)

50. Van Doorn GS. 2009 Intralocus Sexual Conflict. *Ann. N. Y. Acad. Sci.* **1168**, 52–71. (doi:10.1111/j.1749-6632.2009.04573.x)

51. Connallon T, Cox RM, Calsbeek R. 2010 FITNESS CONSEQUENCES OF SEX-SPECIFIC SELECTION. *Evolution* **64**, 1671–1682. (doi:10.1111/j.1558-5646.2009.00934.x)

52. Connallon T, Chenoweth SF. 2019 Dominance reversals and the maintenance of genetic variation for fitness. *PLOS Biol.* **17**, e3000118. (doi:10.1371/journal.pbio.3000118)

53. Connallon T, Clark AG. 2011 The Resolution of Sexual Antagonism by Gene Duplication. *Genetics* **187**, 919–937. (doi:10.1534/genetics.110.123729)

54. Gallach M, Betrán E. 2011 Intralocus sexual conflict resolved through gene duplication. *Trends Ecol. Evol.* **26**, 222–228. (doi:10.1016/j.tree.2011.02.004)

55. Wyman MJ, Cutter AD, Rowe L. 2012 GENE DUPLICATION IN THE EVOLUTION OF SEXUAL DIMORPHISM: DUPLICATES AND SEX-BIASED GENE EXPRESSION. *Evolution* **66**, 1556–1566. (doi:10.1111/j.1558-5646.2011.01525.x)

56. VanKuren NW, Long M. 2018 Gene duplicates resolving sexual conflict rapidly evolved essential gametogenesis functions. *Nat. Ecol. Evol.* **2**, 705–712. (doi:10.1038/s41559-018-0471-0)

57. Charlesworth D, Charlesworth B. 1980 Sex differences in fitness and selection for centric fusions between sex-chromosomes and autosomes. *Genet. Res.* **35**, 205–214. (doi:10.1017/S0016672300014051)

58. Rice WR. 1984 Sex Chromosomes and the Evolution of Sexual Dimorphism. *Evolution* **38**, 735. (doi:10.2307/2408385)

59. Jordan CY, Charlesworth D. 2012 THE POTENTIAL FOR SEXUALLY ANTAGONISTIC POLYMORPHISM IN DIFFERENT GENOME REGIONS. *Evolution* **66**, 505–516. (doi:10.1111/j.1558-5646.2011.01448.x)

60. Bonduriansky R, Day T. 2009 Nongenetic Inheritance and Its Evolutionary Implications. *Annu. Rev. Ecol. Evol. Syst.* **40**, 103–125. (doi:10.1146/annurev.ecolsys.39.110707.173441)

61. Rice WR, Friberg U, Gavrilets S. 2016 Sexually antagonistic epigenetic marks that canalize sexually dimorphic development. *Mol. Ecol.* **25**, 1812–1822. (doi:10.1111/mec.13490)

62. Spencer HG, Priest NK. 2016 The Evolution of Sex-Specific Dominance in Response to Sexually Antagonistic Selection. *Am. Nat.* **187**, 658–666. (doi:10.1086/685827)

63. Keightley PD, Kacser H. 1987 Dominance, pleiotropy and metabolic structure. *Genetics* **117**, 319–329.

64. Hedrick PW. 1999 Antagonistic pleiotropy and genetic polymorphism: a perspective. *Heredity* **82**, 126–133. (doi:10.1038/sj.hdy.6884400)

65. Hoekstra RF, Bijlsma R, Dolman AJ. 1985 Polymorphism from environmental heterogeneity: models are only robust if the heterozygote is close in fitness to the favoured homozygote in each environment. *Genet. Res.* **45**, 299–314. (doi:10.1017/S001667230002228X)

66. Hedrick PW. 1986 Genetic polymorphism in heterogeneous environments: a decade later. *Annu. Rev. Ecol. Syst.* , 535–566.

67. Zouros E, Van Delden W. 1982 Substrate-preference polymorphism at an esterase locus of Drosophila mojavensis. *Genetics* **100**, 307–314.

68. Place AR, Powers DA. 1984 Kinetic characterization of the lactate dehydrogenase (LDH-B4) allozymes of Fundulus heteroclitus. *J. Biol. Chem.* **259**, 1309–1318.

69. MATSUO Y, YAMAZAKI T. 1986 Genetic analysis of natural populations of Drosophila melanogaster in Japan. VI. Differential regulation of duplicated amylase loci and degree of dominance of amylase activity in different environments. *Jpn. J. Genet.* **61**, 543–558.

70. Via S, Bouck AC, Skillman S. 2000 Reproductive isolation between divergent races of pea aphids on two hosts. II. Selection against migrants and hybrids in the parental environments. *Evolution* **54**, 1626–1637.

71. Chen J, Nolte V, Schlötterer C. 2015 Temperature Stress Mediates Decanalization and Dominance of Gene Expression in Drosophila melanogaster. *PLOS Genet.* **11**, e1004883. (doi:10.1371/journal.pgen.1004883)

72. Le Poul Y, Whibley A, Chouteau M, Prunier F, Llaurens V, Joron M. 2014 Evolution of dominance mechanisms at a butterfly mimicry supergene. *Nat. Commun.* **5**, 5644. (doi:10.1038/ncomms6644)

73. Gautier M *et al.* 2018 The Genomic Basis of Color Pattern Polymorphism in the Harlequin Ladybird. *Curr. Biol.* **28**, 3296-3302.e7. (doi:10.1016/j.cub.2018.08.023)

74. Mérot C, Llaurens V, Normandeau E, Bernatchez L, Wellenreuther M. 2020 Balancing selection via life-history trade-offs maintains an inversion polymorphism in a seaweed fly. *Nat. Commun.* **11**, 670. (doi:10.1038/s41467-020-14479-7)

75. Jardine MD, Ruzicka F, Diffley C, Fowler K, Reuter M. 2021 A non-coding indel polymorphism in the *fruitless* gene of *Drosophila melanogaster* exhibits antagonistically pleiotropic fitness effects. *Proc. R. Soc. B Biol. Sci.* **288**, rspb.2020.2958, 20202958. (doi:10.1098/rspb.2020.2958)

76. Pearse DE *et al.* 2019 Sex-dependent dominance maintains migration supergene in rainbow trout. *Nat. Ecol. Evol.* **3**, 1731–1742. (doi:10.1038/s41559-019-1044-6)

77. Geeta Arun M, Agarwala A, Syed ZA, Kashyap M, Venkatesan S, Chechi TS, Gupta V, Prasad NG. 2021 Experimental evolution reveals sex-specific dominance for surviving bacterial infection in laboratory populations of Drosophila melanogaster. *Evol. Lett.* **5**, 657–671.

78. Mishra P, Barrera TS, Grieshop KH, Agrawal AF. 2022 Cis-regulatory variation in relation to sex and sexual dimorphism in Drosophila melanogaster. *bioRxiv*

79. Fisher RA. 1930 *The genetical theory of natural selection: a complete variorum edition*. Oxford University Press.

80. Rutherford A. 2021 Race, eugenics, and the canceling of great scientists. *Am. J. Phys. Anthropol.* **175**, 448–452. (doi:10.1002/ajpa.24192)

81. Orr HA. 2005 The genetic theory of adaptation: a brief history. *Nat. Rev. Genet.* **6**, 119–127. (doi:10.1038/nrg1523)

82. Orr HA. 2005 Theories of adaptation: what they do and don’t say. *Genetica* **123**, 3–13. (doi:10.1007/s10709-004-2702-3)

83. Martin G, Lenormand T. 2006 THE FITNESS EFFECT OF MUTATIONS ACROSS ENVIRONMENTS: A SURVEY IN LIGHT OF FITNESS LANDSCAPE MODELS. *Evolution* **60**, 2413–2427. (doi:10.1111/j.0014-3820.2006.tb01878.x)

84. Connallon T, Clark AG. 2014 Evolutionary inevitability of sexual antagonism. *Proc. R. Soc. B Biol. Sci.* **281**, 20132123. (doi:10.1098/rspb.2013.2123)

85. Connallon T, Clark AG. 2014 Balancing Selection in Species with Separate Sexes: Insights from Fisher’s Geometric Model. *Genetics* **197**, 991–1006. (doi:10.1534/genetics.114.165605)

86. Wright S. 1934 Physiological and Evolutionary Theories of Dominance. *Am. Nat.* **68**, 24–53. (doi:10.1086/280521)

87. Kacser H, Burns JA. 1981 THE MOLECULAR BASIS OF DOMINANCE. *Genetics* **97**, 639–666. (doi:10.1093/genetics/97.3-4.639)

88. Connallon T, Clark AG. 2010 SEX LINKAGE, SEX-SPECIFIC SELECTION, AND THE ROLE OF RECOMBINATION IN THE EVOLUTION OF SEXUALLY DIMORPHIC GENE EXPRESSION: SEX LINKAGE AND SEXUAL DIMORPHISM. *Evolution* **64**, 3417–3442. (doi:10.1111/j.1558-5646.2010.01136.x)

89. Manna F, Martin G, Lenormand T. 2011 Fitness Landscapes: An Alternative Theory for the Dominance of Mutation. *Genetics* **189**, 923–937. (doi:10.1534/genetics.111.132944)

90. Muralidhar P, Veller C. 2022 Dominance shifts increase the likelihood of soft selective sweeps. *Evolution* **76**, 966–984. (doi:10.1111/evo.14459)

91. Dean AM, Dykhuizen DE, Hartl DL. 1986 Fitness as a function of β-galactosidase activity in *Escherichia coli*. *Genet. Res.* **48**, 1–8. (doi:10.1017/S0016672300024587)

92. Dykhuizen DE, Dean AM, Hartl DL. 1987 Metabolic Flux and Fitness. *Genetics* **115**, 25–31. (doi:10.1093/genetics/115.1.25)

93. Dean AM. 1989 Selection and neutrality in lactose operons of Escherichia coli. *Genetics* **123**, 441–454.

94. Papp B, Pál C, Hurst LD. 2003 Dosage sensitivity and the evolution of gene families in yeast. *Nature* **424**, 194–197. (doi:10.1038/nature01771)

95. Dekel E, Alon U. 2005 Optimality and evolutionary tuning of the expression level of a protein. *Nature* **436**, 588–592. (doi:10.1038/nature03842)

96. Phadnis N, Fry JD. 2005 Widespread Correlations Between Dominance and Homozygous Effects of Mutations: Implications for Theories of Dominance. *Genetics* **171**, 385–392. (doi:10.1534/genetics.104.039016)

97. Zeyl C, Vanderford T, Carter M. 2003 An Evolutionary Advantage of Haploidy in Large Yeast Populations. *Science* **299**, 555–558. (doi:10.1126/science.1078417)

98. Agrawal AF, Whitlock MC. 2011 Inferences About the Distribution of Dominance Drawn From Yeast Gene Knockout Data. *Genetics* **187**, 553–566. (doi:10.1534/genetics.110.124560)

99. Meisel RP, Connallon T. 2013 The faster-X effect: integrating theory and data. *Trends Genet.* **29**, 537–544. (doi:10.1016/j.tig.2013.05.009)

100. Ronfort J, Glémin S. 2013 Mating system, Haldane's sieve, and the domestication process. *Evolution* **67**, 1518–1526.

101. Fry JD. 2009 THE GENOMIC LOCATION OF SEXUALLY ANTAGONISTIC VARIATION: SOME CAUTIONARY COMMENTS. *Evolution* (doi:10.1111/j.1558-5646.2009.00898.x)

102. Agrawal AF. 2009 Spatial Heterogeneity and the Evolution of Sex in Diploids. *Am. Nat.* **174**, S54–S70. (doi:10.1086/599082)

103. Reid JM. 2022 Intrinsic emergence and modulation of sex-specific dominance reversals in threshold traits. *Evolution*

104. Otto SP, Bourguet D. 1999 Balanced Polymorphisms and the Evolution of Dominance. *Am. Nat.* **153**, 561–574. (doi:10.1086/303204)

105. Provine WB. 1992 The R. A. Fisher—Sewall Wright Controversy. In *The Founders of Evolutionary Genetics* (ed S Sarkar), pp. 201–229. Dordrecht: Springer Netherlands. (doi:10.1007/978-94-011-2856-8_8)

106. Mayo O, Bürger R. 1997 THE EVOLUTION OF DOMINANCE: A THEORY WHOSE TIME HAS PASSED? *Biol. Rev. Camb. Philos. Soc.* **72**, 97–110. (doi:10.1017/S0006323196004987)

107. Flintham E, Savolainen V, Otto S, Reuter M, Mullon C. 2023 The maintenance of genetic polymorphism in sexually antagonistic traits. , 2023.10.10.561678. (doi:10.1101/2023.10.10.561678)

108. Friedlander T, Prizak R, Barton NH, Tkačik G. 2017 Evolution of new regulatory functions on biophysically realistic fitness landscapes. *Nat. Commun.* **8**, 216. (doi:10.1038/s41467-017-00238-8)

109. Porter AH, Johnson NA, Tulchinsky AY. 2017 A New Mechanism for Mendelian Dominance in Regulatory Genetic Pathways: Competitive Binding by Transcription Factors. *Genetics* **205**, 101–112. (doi:10.1534/genetics.116.195255)

110. Wong HWS, Holman L. 2023 Pleiotropic fitness effects across sexes and ages in the Drosophila genome and transcriptome. *Evolution* **77**, 2642–2655. (doi:10.1093/evolut/qpad163)

111. Prout T. 2000 How well does opposing selection maintain variation? In *Evolutionary Genetics* (eds RS Singh, CB Krimbas), pp. 157–181. Cambridge: Cambridge University Press.

112. Wittmann MJ, Bergland AO, Feldman MW, Schmidt PS, Petrov DA. 2017 Seasonally fluctuating selection can maintain polymorphism at many loci via segregation lift. *Proc. Natl. Acad. Sci.* **114**, E9932–E9941. (doi:10.1073/pnas.1702994114)

113. Bertram J, Masel J. 2019 Different mechanisms drive the maintenance of polymorphism at loci subject to strong versus weak fluctuating selection. *Evolution* **73**, 883–896. (doi:10.1111/evo.13719)

114. Lynch M, Blanchard J, Houle D, Kibota T, Schultz S, Vassilieva L, Willis J. 1999 PERSPECTIVE: SPONTANEOUS DELETERIOUS MUTATION. *Evolution* **53**, 645–663. (doi:10.1111/j.1558-5646.1999.tb05361.x)

115. Lynch M, Walsh B. 1998 Genetics and analysis of quantitative traits.

116. Su W, Michaud JP, Runzhi Z, Fan Z, Shuang L. 2009 Seasonal cycles of assortative mating and reproductive behaviour in polymorphic populations of Harmonia axyridis in China. *Ecol. Entomol.* **34**, 483–494.

117. Czorlich Y, Aykanat T, Erkinaro J, Orell P, Primmer CR. 2018 Rapid sex-specific evolution of age at maturity is shaped by genetic architecture in Atlantic salmon. *Nat. Ecol. Evol.* **2**, 1800–1807.

118. Aykanat T, Ozerov M, Vähä J-P, Orell P, Niemelä E, Erkinaro J, Primmer CR. 2019 Co-inheritance of sea age at maturity and iteroparity in the Atlantic salmon vgll3 genomic region. *J. Evol. Biol.* **32**, 343–355.

119. Hayman BI. 1954 The Theory and Analysis of Diallel Crosses. *Genetics* **39**, 789–809.

120. Kendall NW, McMillan JR, Sloat MR, Buehrens TW, Quinn TP, Pess GR, Kuzishchin KV, McClure MM, Zabel RW. 2015 Anadromy and residency in steelhead and rainbow trout (Oncorhynchus mykiss): a review of the processes and patterns. *Can. J. Fish. Aquat. Sci.* **72**, 319–342.

121. Ohms HA, Sloat MR, Reeves GH, Jordan CE, Dunham JB. 2014 Influence of sex, migration distance, and latitude on life history expression in steelhead and rainbow trout (Oncorhynchus mykiss). *Can. J. Fish. Aquat. Sci.* **71**, 70–80.

122. Kasimatis KR, Nelson TC, Phillips PC. 2017 Genomic Signatures of Sexual Conflict. *J. Hered.* **108**, 780–790. (doi:10.1093/jhered/esx080)

123. Kasimatis KR, Ralph PL, Phillips PC. 2019 Limits to Genomic Divergence Under Sexually Antagonistic Selection. *G3 GenesGenomesGenetics* **9**, 3813–3824. (doi:10.1534/g3.119.400711)

124. Rowe L, Chenoweth SF, Agrawal AF. 2018 The Genomics of Sexual Conflict. *Am. Nat.* **192**, 274–286. (doi:10.1086/698198)

125. Ruzicka F *et al.* 2020 The search for sexually antagonistic genes: Practical insights from studies of local adaptation and statistical genomics. *Evol. Lett.* **4**, 398–415. (doi:10.1002/evl3.192)

126. Kasimatis KR, Abraham A, Ralph PL, Kern AD, Capra JA, Phillips PC. 2021 Evaluating human autosomal loci for sexually antagonistic viability selection in two large biobanks. *Genetics* **217**, iyaa015. (doi:10.1093/genetics/iyaa015)

127. Connallon T, Matthews G. 2019 Cross-sex genetic correlations for fitness and fitness components: Connecting theoretical predictions to empirical patterns. *Evol. Lett.* **3**, 254–262. (doi:10.1002/evl3.116)

128. Bergland AO, Behrman EL, O'Brien KR, Schmidt PS, Petrov DA. 2014 Genomic Evidence of Rapid and Stable Adaptive Oscillations over Seasonal Time Scales in Drosophila. *PLoS Genet.* **10**, e1004775. (doi:10.1371/journal.pgen.1004775)

129. Buffalo V, Coop G. 2020 Estimating the genome-wide contribution of selection to temporal allele frequency change. *Proc. Natl. Acad. Sci.* **117**, 20672–20680. (doi:10.1073/pnas.1919039117)

130. Machado HE *et al.* 2021 Broad geographic sampling reveals the shared basis and environmental correlates of seasonal adaptation in Drosophila. *eLife* **10**, e67577. (doi:10.7554/eLife.67577)

131. Singh A, Hasan A, Agrawal AF. 2023 An investigation of the sex-specific genetic architecture of fitness in Drosophila melanogaster. *Evolution* **77**, 2015–2028. (doi:10.1093/evolut/qpad107)

132. Gerstein AC, Kuzmin A, Otto SP. 2014 Loss-of-heterozygosity facilitates passage through Haldane's sieve for Saccharomyces cerevisiae undergoing adaptation. *Nat. Commun.* **5**, 3819. (doi:10.1038/ncomms4819)

133. Palmer DS *et al.* 2023 Analysis of genetic dominance in the UK Biobank. *Science* **379**, 1341–1348. (doi:10.1126/science.abn8455)

134. Signor SA, Nuzhdin SV. 2018 The Evolution of Gene Expression in cis and trans. *Trends Genet.* **34**, 532–544. (doi:10.1016/j.tig.2018.03.007)

135. Rhoads A, Au KF. 2015 PacBio Sequencing and Its Applications. *Genomics Proteomics Bioinformatics* **13**, 278–289. (doi:10.1016/j.gpb.2015.08.002)

136. Jain M, Olsen HE, Paten B, Akeson M. 2016 The Oxford Nanopore MinION: delivery of nanopore sequencing to the genomics community. *Genome Biol.* **17**, 239. (doi:10.1186/s13059-016-1103-0)

137. Kasimatis KR, Sánchez-Ramírez S, Stevenson ZC. 2021 Sexual Dimorphism through the Lens of Genome Manipulation, Forward Genetics, and Spatiotemporal Sequencing. *Genome Biol. Evol.* **13**, evaa243. (doi:10.1093/gbe/evaa243)

138. Naftaly AS, Pau S, White MA. 2021 Long-read RNA sequencing reveals widespread sex-specific alternative splicing in threespine stickleback fish. *Genome Res.* , gr. 274282.120.

139. Glinos DA *et al.* 2022 Transcriptome variation in human tissues revealed by long-read sequencing. *Nature* **608**, 353–359.

140. Tilgner H, Jahanbani F, Gupta I, Collier P, Wei E, Rasmussen M, Snyder M. 2018 Microfluidic isoform sequencing shows widespread splicing coordination in the human transcriptome. *Genome Res.* **28**, 231–242. (doi:10.1101/gr.230516.117)

141. Kulkarni A, Anderson AG, Merullo DP, Konopka G. 2019 Beyond bulk: a review of single cell transcriptomics methodologies and applications. *Curr. Opin. Biotechnol.* **58**, 129–136. (doi:10.1016/j.copbio.2019.03.001)

142. Buenrostro JD, Wu B, Chang HY, Greenleaf WJ. 2015 ATAC-seq: A Method for Assaying Chromatin Accessibility Genome-Wide. *Curr. Protoc. Mol. Biol.* **109**. (doi:10.1002/0471142727.mb2129s109)

143. Yan F, Powell DR, Curtis DJ, Wong NC. 2020 From reads to insight: a hitchhiker's guide to ATAC-seq data analysis. *Genome Biol.* **21**, 1–16.

144. Chen X, Litzenburger UM, Wei Y, Schep AN, LaGory EL, Choudhry H, Giaccia AJ, Greenleaf WJ, Chang HY. 2018 Joint single-cell DNA accessibility and protein epitope profiling reveals environmental regulation of epigenomic heterogeneity. *Nat. Commun.* **9**, 1–12.

145. Xu S *et al.* 2020 regSNPs-ASB: A Computational Framework for Identifying Allele-Specific Transcription Factor Binding From ATAC-seq Data. *Front. Bioeng. Biotechnol.* **8**, 886. (doi:10.3389/fbioe.2020.00886)

146. Bentsen M *et al.* 2020 ATAC-seq footprinting unravels kinetics of transcription factor binding during zygotic genome activation. *Nat. Commun.* **11**, 4267. (doi:10.1038/s41467-020-18035-1)

**ELECTRONIC SUPPLEMENTARY MATERIAL**

# Dominance reversals: The resolution of genetic conflict and maintenance of genetic variation

Karl Grieshop[1,2,3*], Eddie K. H. Ho[4] and Katja R. Kasimatis[1,5]

[1]Department of Ecology and Evolutionary Biology, University of Toronto, Toronto, Canada

[2]Department of Molecular Biosciences, The Wenner-Gren Institute, Stockholm University, Stockholm, Sweden

[3]School of Biological Sciences, University of East Anglia, Norwich Research Park, Norwich, NR4 7TJ, United Kingdom

[4]Department of Biology, Reed College, Portland, Oregon, United States

[5]Department of Biology, University of Virginia, Charlottesville, Virginia, United States

ORCIDs:

https://orcid.org/0000-0001-8925-5066 (KG)

https://orcid.org/0000-0002-7692-394X (EKHH)

https://orcid.org/0000-0003-0025-5022 (KRK)

*Corresponding author:

Email: K.Grieshop@uea.ac.uk (KG)

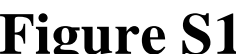

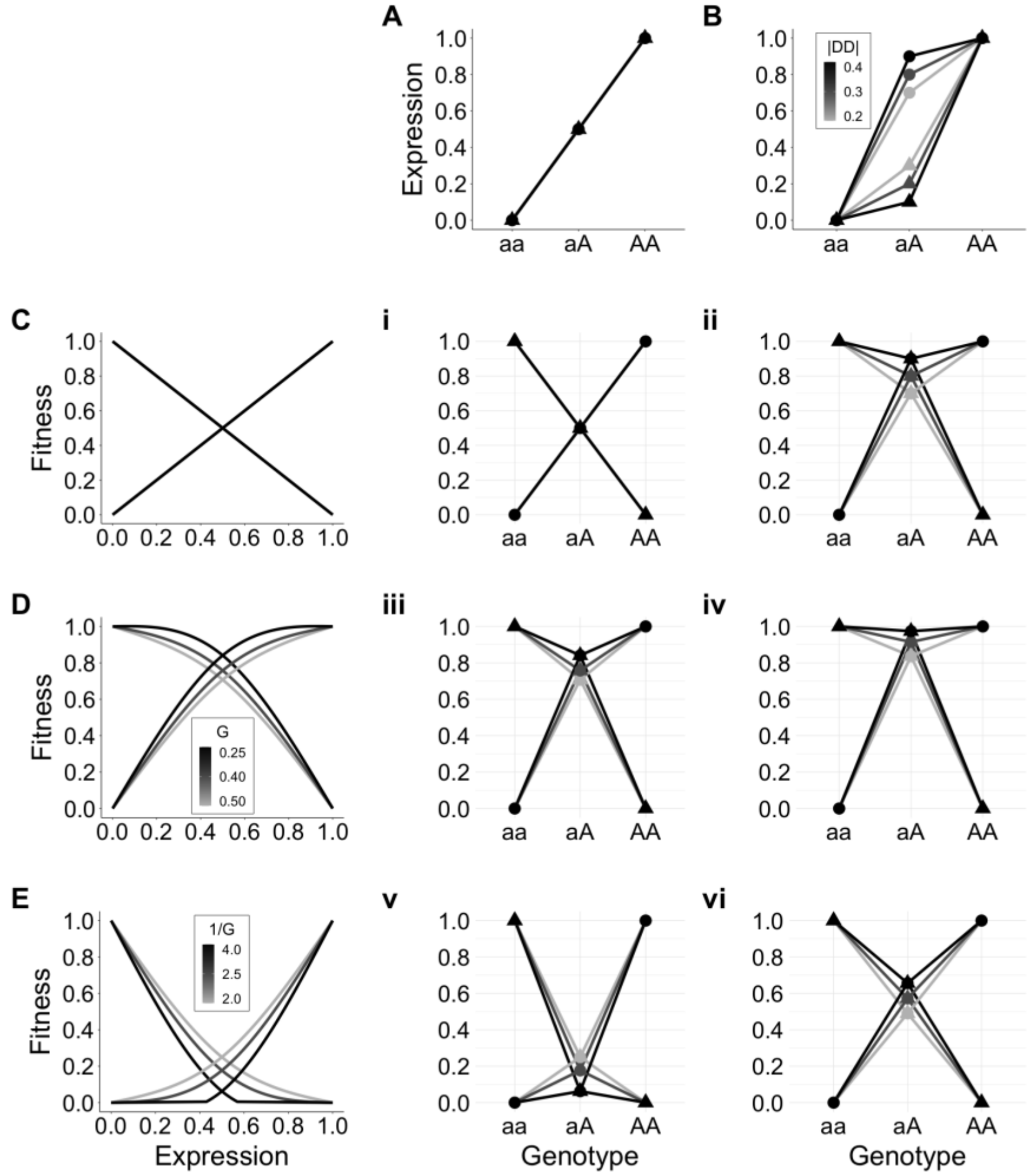

**Figure S1. Graphical representation of dominance reversals.** Circles versus triangles represent alternative sexes, environments, or generations (antagonistic pleiotropy scenarios not shown). Grey to black shading represents different strengths of an effect (|DD| = absolute value dominance deviation from additivity (B); G = the exponential scaling of the fitness landscape (D)). We first map genotype to trait expression under two scenarios: **(A)** additivity for the phenotype, and **(B)** dominance reversal for the phenotype. Then we map trait expression to fitness under two scenarios of antagonistic selection: **(C)** additivity for fitness and **(D)** curved fitness functions overlapping in their concave vicinities. The ultimate effect of a genotype on fitness is given by the combination of the genotype-phenotype map (**A or B**) and the phenotype-fitness map (**C or D**) in the matrix of resultant genotype-fitness panels **i-iv**: **(i)** additivity, **(ii)** dominance reversal owing to the genotype-phenotype map (**ii**) or owing to the phenotype-fitness map (**iii**), a greater magnitude of dominance reversal owing to their combined effects **(iv)**, net heterozygote inferiority owing to the fitness landscape **(v)**, and various outcomes depending on the magnitudes of |DD| and G that are combined (e.g. canceling out to apparent additivity) **(vi)**.

**S2. Simulating the evolution of a biophysically explicit dominance modifier**

**S2. Methods**

Our simulation model builds off of Porter et al.[1] and is based on the premise that transcription factors (TF) and binding sites behave according to the thermodynamic and kinetic properties of molecular interactions. The proportion of time that a binding site is occupied by a TF molecule (fractional occupancy) determines gene expression levels. For a generic diploid locus where TF variants $T_1$ and $T_2$ compete for occupancy at promotor (or repressor) sites $P_1$ and $P_2$, we utilize equation (2a) from Porter et al.[1] to define the fractional occupancy of $T_1$ on $P_1$ as:

$$F([T_1],[T_2],\ m_{11}, m_{21},\ k) = \frac{\frac{[T_1]}{k^{m_{11}}}}{1+\frac{[T_1]}{k^{m_{11}}}+\frac{[T_2]}{k^{m_{21}}}}. \tag{1.1}$$

The square brackets denote the concentration of the TF, e.g., $[T_1]$ represents the concentration of $T_1$. $m_{11}$ represents the proportion of mismatched bits between $T_1$ and $P_1$, $m_{21}$ presents the proportion of mismatched bits between $T_2$ and $P_1$. Lastly, $k$ represents the stepwise change in the dissociation constant[1].

Our diploid model consists of two unlinked interacting genes A and B (Figure S2.1, Table S1.1). Our model is likely an appropriate simplification of reality: while real gene regulation may be considerably more complicated at a minimum[2], that may be largely due to molecular redundancies owing to the evolutionary entrenchment of gene networks down paths of no return[3]. We will use the subscript $i = \{1, 2\}$ to represent the homolog to which the gene belongs. Gene A contains a *cis*-regulatory binding site represented by $\alpha_i$ which possesses 11 allelic variants with values between 0 and 1 in 0.1 increments; $\alpha_i = \{0, 0.1, \dots, 0.9, 1\}$. These 11 different functional alleles mimic the possible states of a typical eukaryotic transcription factor binding site – from all 10 nucleotides matching to none matching[4]. Gene A also contains a bi-allelic coding region with variants $A_i = \{1, 2\}$, which expresses a TF that regulates the expression of Gene B. Importantly, the TF expressed by allele $A_i = 1$ ultimately downregulates the expression of Gene B, while the TF expressed by allele $A_i = 2$ ultimately upregulates the expression of Gene B (see equation (2.6), below). Expression of $A_i$ ($[A_i]$) is controlled by the binding of an anticorrelated, sex-limited regulatory stimulus, $D$, onto the binding site $\alpha_i$. This sex-limited regulatory stimulus could be thought of as approximating a sex-limited hormone, or alternatively spliced *dsx* or *fru* genes in insects[5]. The fractional occupancy of $D$, onto the binding site $\alpha_i$ depends on its concentration, $[D]$, the proportion of mismatched bits, $m_{D\alpha}$, and the stepwise change in the dissociation constant $k_\alpha$ (Porter et al.[1]). Importantly, $m_{D\alpha}$ depends on the sex of the individual and the allelic variant of $\alpha_i$. The male version of the regulatory stimuli, *D*, binds optimally to allele $\alpha_i = 0$, and the female version of *D* binds optimally to allele $\alpha_i = 1$:

$$m_{D\alpha}(sex,\ \alpha_i) = \begin{cases} \alpha_i, & sex = male \\ 1 - \alpha_i, & sex = female \end{cases}. \tag{1.2}$$

Thus, $[A_i]$ is the product of the fractional occupancy and $G_A$, which represents the maximum concentration that can be achieved, so that the concentration of $A_1$ and $A_2$ is calculated as follows (Figure S2.1, Table S1.1):

$$[A_1] = G_A * \left(\frac{[D]/{k_\alpha}^{m_{D\alpha}(sex,\alpha_1)}}{1+2[D]/{k_\alpha}^{m_{D\alpha}(sex,\alpha_1)}}\right), \tag{1.3}$$

$$[A_2] = G_A * \left(\frac{[D]/{k_\alpha}^{m_{D\alpha}(sex,\alpha_2)}}{1+2[D]/{k_\alpha}^{m_{D\alpha}(sex,\alpha_2)}}\right). \tag{1.4}$$

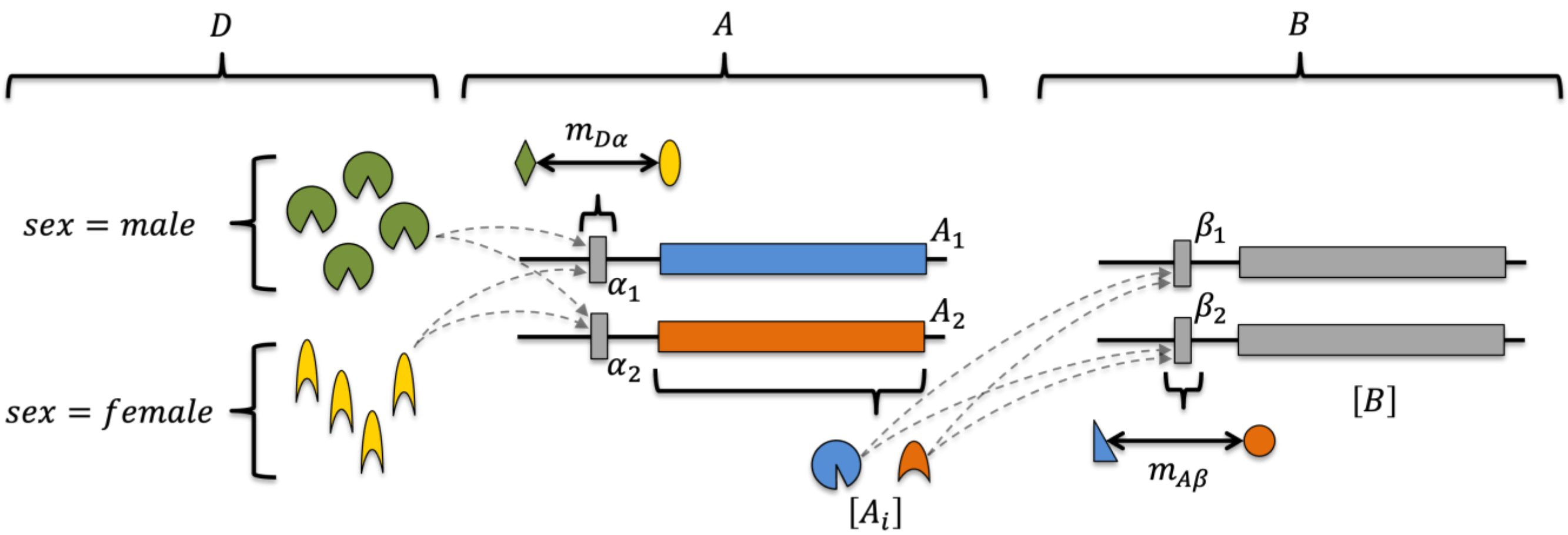

**Figure S2.1: Cartoon schematic of some of the parameters defined in Table S1.1.** Due to the sex-specific fitness optima for $\varphi$ (the standardized expression of Gene B, $[B]$, see below), and the fact that $A_i = 1$ and $A_i = 2$ affect $[B]$ in opposite directions, $A_i = 1$ and $A_i = 2$ represent the male- and female-benefit alleles, respectively, of the focal sexually antagonistic polymorphism at $A_i$. $D$, $A$ and $B$ are unlinked, whereas the protein-coding sites of Genes A and B, $A_i$ and $B_i$, are linked to their respective *cis*-regulatory binding sites, $\alpha_i$ and $\beta_i$.

**Table S2.1**

| Parameter | Definition |
|---|---|
| $i$ | Index for homolog; $i = \{1, 2\}$. |
| $\alpha_i$ | Allelic state of $\alpha$ at homologue $i$; $\alpha_i = \{0, 0.1, \dots 0.9, 1\}$. |
| $A_i$ | Allelic state of $A$ at homologue $i$; $A_i = 1$ reduces $[B]$, and $A_i = 2$ elevates $[B]$. |
| $\beta_i$ | Allelic state of $\beta$ at homologue $i$; $\beta_i = \{0, 0.1, \dots 0.9, 1\}$. |
| $m_{D\alpha}(sex, \alpha_i)$ | Proportion of mismatched bits between $D$ and $\alpha_i$ as a function of sex and $\alpha_i$. |
| $m_{A\beta}(A_i, \beta_i)$ | Proportion of mismatched bits between $A_i$ and $\beta_i$. |
| $k_\alpha$ | Stepwise change in the dissociation constant for $D$ binding to $\alpha_i$. |
| $k_\beta$ | Stepwise change in the dissociation constant for $A_i$ binding to $\beta_i$. |
| $G_A$ | Concentration for protein of allele $A_i$ if fractional occupancy at binding site is 1. $G_A \leq [A_{sat}]$. |
| $H(A_i)$ | Effect of $A_i$ on $[B]$; $H(A_i) = \begin{cases} -1, & A_i = 1 \\ +1, & A_i = 2 \end{cases}$. |
| $[D]$ | Concentration of sex-limited regulatory stimulus. |
| $[A_i]$ | Concentration of protein expressed by $A_i$. |
| $[B]$ | Concentration (expression level) of protein expressed by $B$. |
| $Z$ | Baseline $[B]$. |
| $[A_{sat}]$ | $[A_i]$ that would saturate binding sites of $B$; determines $[B_{\min}]$ and $[B_{\max}]$. |
| $[B_{\min}]$ | $[B]$ when saturated by $A_i = 1$, which reduces expression. |
| $[B_{\max}]$ | $[B]$ when saturated by $A_2 = 2$, which elevates expression. |
| $\varphi$ | Standardized $[B]$. |
| $s_m, s_f$ | Selection coefficient for males and females, respectively. |
| $\gamma_m, \gamma_f$ | Fitness curvature for males and females, respectively. |
| $W_m(\varphi), W_f(\varphi)$ | Absolute fitness for males and females, respectively. |

Gene B contains a *cis*-regulatory binding site represented by $\beta_i$ which possesses 11 allelic variants ($\beta_i = \{0, 0.1, \dots, 0.9, 1\}$) and a monomorphic coding region from which the expression level, $[B]$, ultimately determines the fitness of the individual. The fractional occupancy of the TF expressed by $A_i$, onto the binding site $\beta_i$ depends on its concentration, $[A_i]$, the proportion of mismatched bits, $m_{A\beta}$, and the stepwise change in the dissociation constant $k_\beta$. $m_{A\beta}$ depends on the allelic state of both $A_i$ and $\beta_i$, such that allele $A_i = 1$ binds optimally to allele $\beta_i = 0$, and allele $A_i = 2$ binds optimally to allele $\beta_i = 1$:

$$m_{A\beta}(A_i,\ \beta_i) = \begin{cases} \beta_i, & A_i = 1 \\ 1 - \beta_i, & A_i = 2 \end{cases}. \tag{1.5}$$

The expression level of Gene B, $[B]$, is determined by the combination of all four possible allele-specific interactions between $[A_i]$ and $\beta_i$ like so:

$$\begin{gathered} [B] = Z + \\ \tfrac{1}{2}\Big(F\big([A_1], [A_2], m_{A\beta}(A_1, \beta_1), m_{A\beta}(A_2, \beta_1), k_\beta\big)H(A_1) + \\ F\big([A_2], [A_1], m_{A\beta}(A_2, \beta_1), m_{A\beta}(A_1, \beta_1), k_\beta\big)H(A_2) + \\ F\big([A_1], [A_2], m_{A\beta}(A_1, \beta_2), m_{A\beta}(A_2, \beta_2), k_\beta\big)H(A_1) + \\ F([A_2], [A_2], m_{A\beta}(A_2, \beta_2), m_{A\beta}(A_1, \beta_2), k_\beta)H(A_2)\Big), \\ H(A_i) = \begin{cases} -1, & A_i = 1 \\ +1, & A_i = 2 \end{cases}, \end{gathered} \tag{1.6}$$

where $Z$ represents the base line expression level of Gene B, and $H$ scales the direction of the allele-specific effects of $A_i$ on $[B]$ such that $A_i = 1$ downregulates and $A_i = 2$ upregulates. When an individual is homozygous at the $A_i$ locus (i.e., $A_1 = A_2$), the first two terms in equation 1.6 become equivalent and the last two terms in equation 1.6 also become equivalent. We normalize $[B]$ to the range [0, 1] using:

$$\varphi = \frac{[B] - [B_{min}]}{[B_{max}] - [B_{min}]}. \tag{1.7}$$

The minimum $[B]$ occurs when $A_1 = A_2 = 1$, $[A_1] + [A_2] = [A_{sat}]$ (see Table S1.1), and $\beta_i = 0$ to yield $[B_{min}] = Z - [A_{sat}]/(1 + [A_{sat}])$. The maximum $[B]$ occurs when $A_1 = A_2 = 2$, $[A_1] + [A_2] = [A_{sat}]$, and $\beta_i = 1$ to yield $[B_{max}] = Z + [A_{sat}]/(1 + [A_{sat}])$.

Male- and female-specific fitness as a function of the normalized expression, φ, is represented by $W_m(\varphi)$ and $W_f(\varphi)$, respectively, calculated as

$$W_m(\varphi) = 1 - (s_m\varphi)^{\gamma_m} \tag{1.8}$$

and

$$W_f(\varphi) = 1 - (s(1-\varphi))^{\gamma_f}, \tag{1.9}$$

where $s_m$ and $s_f$ are the sex-specific selection coefficients, and $\gamma_m$ and $\gamma_f$ control the curvature of the fitness landscape for each sex (e.g., $\gamma_m = 1$ for linear/additive fitness effects). The opposite sign of sex-specific fitness consequences for φ represents sex-specific fitness optima, and because alternative alleles $A_i = 1$ and $A_i = 2$ affect φ in opposite directions (via their effects on $[B]$, see equation (1.6) and Table S1.1), they respectively represent the focal male- and female-benefit alleles of a sexually antagonistic polymorphism (Figure S2.1).

We used simulations to assess whether this model can maintain polymorphism at $A_i$. All simulations consisted of 5,000 male and 5,000 female individuals, and ran for 50,000 non-overlapping generations. At the start of the simulation, for each individual, the values of each allele at each locus ($A_i$, $\alpha_i$, and $\beta_i$) were randomly selected with equal probability from their corresponding set of possible values; recall that $A_i$ can have values $\{1, 2\}$, while $\alpha_i$, and $\beta_i$ can have values $\{0, 0.1, 0.2, \ldots, 0.9, 1\}$. Every generation, each of the 10,000 offspring were produced as follows. A potential female parent is chosen randomly and their fitness relative to the maximum among females, $W_f(\varphi) \,/\, W_{f,max}$, was compared to a uniform random deviate (Uniform [0, 1]), where $W_{f,max}$ is the largest value of $W_f(\varphi)$ in the population. The female was kept if their relative fitness surpasses the uniform random deviate, otherwise, the process was repeated until a suitable female was chosen. A male was chosen using the same procedure. Gametes from each parent were produced following segregation between the unlinked Genes A and B. After combining the two gametes from the parents, the alleles at $\alpha_i$ and $\beta_i$ had a probability $\mu_\alpha$ and $\mu_\beta$, respectively, of mutating to an adjacent value. If $\alpha_i$ or $\beta_i$ has allelic value 0 then it could only mutate to 0.1 and analogously, if $\alpha_i$ or $\beta_i$ has allelic value 1 then it could only mutate to 0.9. Finally, 5,000 offspring were randomly assigned to be females and the remaining 5,000 assigned to be males.

We present simulations run with a range of selection coefficients, $s_f = s_m = \{0.005, 0.01, 0.02, 0.05\}$. We ran all simulations under a linear fitness landscape, $\gamma_f = \gamma_m = 1$, as we were interested in whether this gene regulatory network could adaptively evolve to constitute a sex-specific "dominance modifier" that resolves genetic conflict and protects the focal polymorphisms from becoming fixed *without* additional beneficial reversal of dominance stemming from the concavity of the fitness landscape. One set of simulations allowed both *cis*-regulatory binding sites on Gene A and B ($\alpha_i$ and $\beta_i$) to mutate with rates $\mu_\alpha = \mu_\beta = 0.005$. Another set of simulations only allowed only $\alpha_i$ to mutate with $\mu_\alpha = 0.005$ and fixed the value of $\beta_i$ at 0.5 for all individuals ($\mu_\beta = 0$). The contrast between simulations that allow $\beta_i$ to mutate/evolve and those that do not serves as an approximation of the role of the epistatic interaction between Genes A and B in influencing the evolution and signature of the sex-specific dominance modifier, while keeping all else equal. Each parameter set was replicated 100 times and the data from the last generation (i.e., 50,000) was used for analysis.

We focus our analysis on the polymorphism at the bi-allelic coding region of Gene A. Since $A_i$ does not undergo mutation, simulations either end with the population having maintained polymorphism at $A_i$ or fixed for one of the two alleles $A_i = 1$ or $A_i = 2$. Signatures used to assess whether the polymorphism at $A_i$ was maintained by sexually antagonistic selection (as opposed to drift) included: the presence of beneficial reversals of dominance for fitness between alternative alleles at $A_i$, the relative abundance of high-fitness $\alpha_i A_i$ haplotypes, the average distance to the sex-specific fitness optima, and the sex-specific allelic imbalance at $A_i$. We considered there to be beneficial dominance reversal when the harmonic mean fitness of individuals heterozygous at the $A_i$ locus (genotype denoted as 1/2) is higher than the harmonic mean fitnesses of homozygous 1/1 genotypes and of homozygous 2/2 genotypes. We use the correlation between the allelic values at $\alpha_i$ and $A_i$ to gauge the presence of high-fitness $\alpha_i A_i$ haplotypes. A positive correlation suggests that allele $A_i = 1$ tend to be linked to $\alpha_i$ alleles with small values (i.e., male benefit haplotypes) and that allele $A_i = 2$ tend to be linked to $\alpha_i$ alleles with large values (i.e., female benefit haplotypes). The average distance to the fitness optima for males was calculated as the normalized expression level of Gene B, φ, averaged across all males. The average distance to fitness optima for females was calculated as 1 minus φ, averaged across all females. Allelic imbalance for individuals heterozygous at the $A_i$ locus was calculated as the $\log_2$ ratio of $[A_i = 2]$ to $[A_j = 1]$, such that positive values indicate the expression of allele $A_i = 2$ is higher than that for $A_i = 1$, and negative values indicate the opposite.

In addition, we interpret a unimodal distribution of $\alpha$ or $\beta$ alleles at the end state as representing the fixation of those sites, and bimodal distribution of $\alpha$ or $\beta$ alleles at the end state as representing maintenance of a functionally bi-allelic polymorphism. We anticipate that these unimodal versus bimodal alternatives would culminate in even more discrete, truly monomorphic or bi-allelic, equilibria under even more realistic parameter settings (namely, very low mutation rates and a very long number of generations). While the parameter settings for mutation rates in evolutionary simulations are typically arbitrary due to computational and time limits, we note that one consequence of setting a mutation rate for what is meant to reflect a 10-nucleotide sequence is that the per-nucleotide mutation rates that we are approximating are actually an order of magnitude lower than the value of our parameter setting. That is, to set $\mu_\alpha = 0.005$, for example, is to model a per-nucleotide mutation rate of 0.0005 for $\alpha$, since the mismatch $m_{D\alpha}$ of an allele at $\alpha_i$ approximates the total number of (mis)matched nucleotides in a binding site, which consists of 10 independently mutating nucleotides. This understanding is only trivially important if/when considering the relationship between the strengths of mutation and selection upon interpreting our results.

## S2 Results

Whether or not polymorphism at $A_i$ is maintained by the action of the surrounding gene regulatory phenomena is sensitive to the parameter settings chosen (Table S2.2, Figure S2.2). In general, $s_m$ and $s_f$ must be sufficiently stronger than $\mu_\alpha$ (as well as $\mu_\beta$, when allowed) in order for selection to maintain the focal polymorphism at $A_i$.

**Table S1.2. Proportion of simulations that ended with polymorphism at locus $A_i$**

| Selection strength | Binding site mutation rates | |
|---|---|---|
| | $\mu_\alpha = \mu_\beta = 0.005$ | $\mu_\alpha = 0.005, \mu_\beta = 0$ |
| 0.005 | 0.2 | 0.16 |
| 0.01 | 0.37 | 0.25 |
| 0.02 | 0.8 | 0.65 |
| 0.05 | 0.95 | 0.99 |

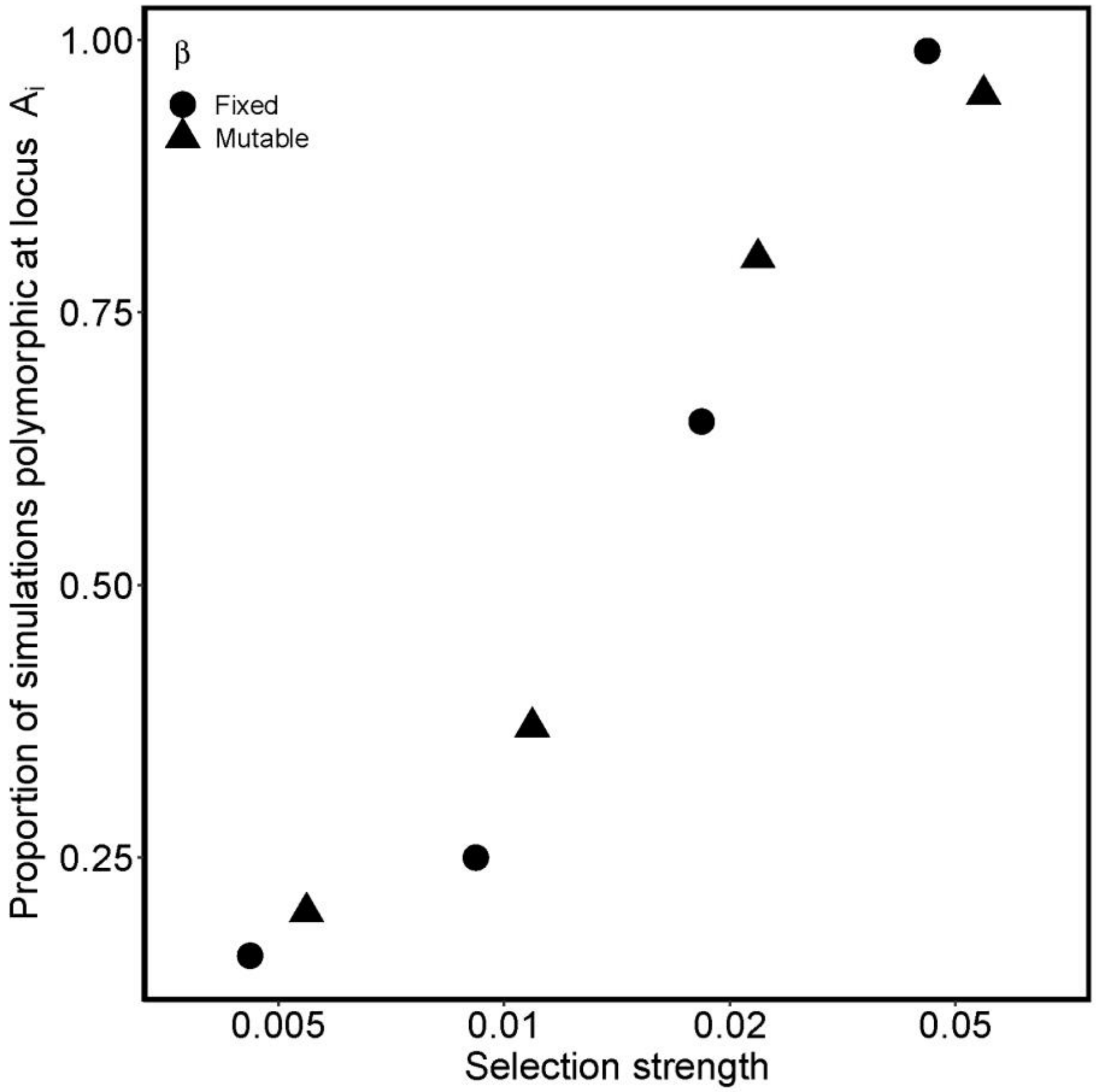

**Figure S2.2: Proportion of 100 simulations maintaining polymorphism at $A_i$ for a range of selection strengths.** Circles and triangles represent simulations in which $\beta_i$ was fixed and mutable, respectively.

Provided the focal polymorphism was maintained, three lines of reasoning suggest that it is due to the surrounding gene regulatory phenomena enabling selection to generate a dominance reversal that partially resolves the genetic conflict. First, the proportion of simulations that resulted in a beneficial reversal of dominance increased with increasing strength of selection (Figure S2.3). Second, there was typically a high prevalence of the most adaptive haplotypes relative to the non-adaptive haplotypes for Gene A, measured as a high correlation between the allelic values at $\alpha_i$ and $A_i$ (Figure S2.4). The positive correlation indicates that the male-benefit allele, $A_i = 1$, tends to be linked to $\alpha_i$ alleles with greater binding affinity to the male-limited regulatory stimulus, and that the female-benefit allele, $A_i = 2$, tends to be linked with $\alpha_i$ alleles with a greater binding affinity to the female-limited regulatory stimulus. Third, the average distance of male and female individuals to their respective fitness optima was less than it would be for a simple codominant polymorphism (i.e., < 0.5) (Figure S2.5). This adaptive sex-specific dominance reversal yields a characteristic pattern of allele-specific expression akin to the findings of Chen et al.[6] and Mishra et al.[7], in which $A_i = 2$ is expressed higher than $A_i = 1$ in female heterozygotes (positive values of allelic imbalance), and *vice versa* for male heterozygotes (Figure S2.6).

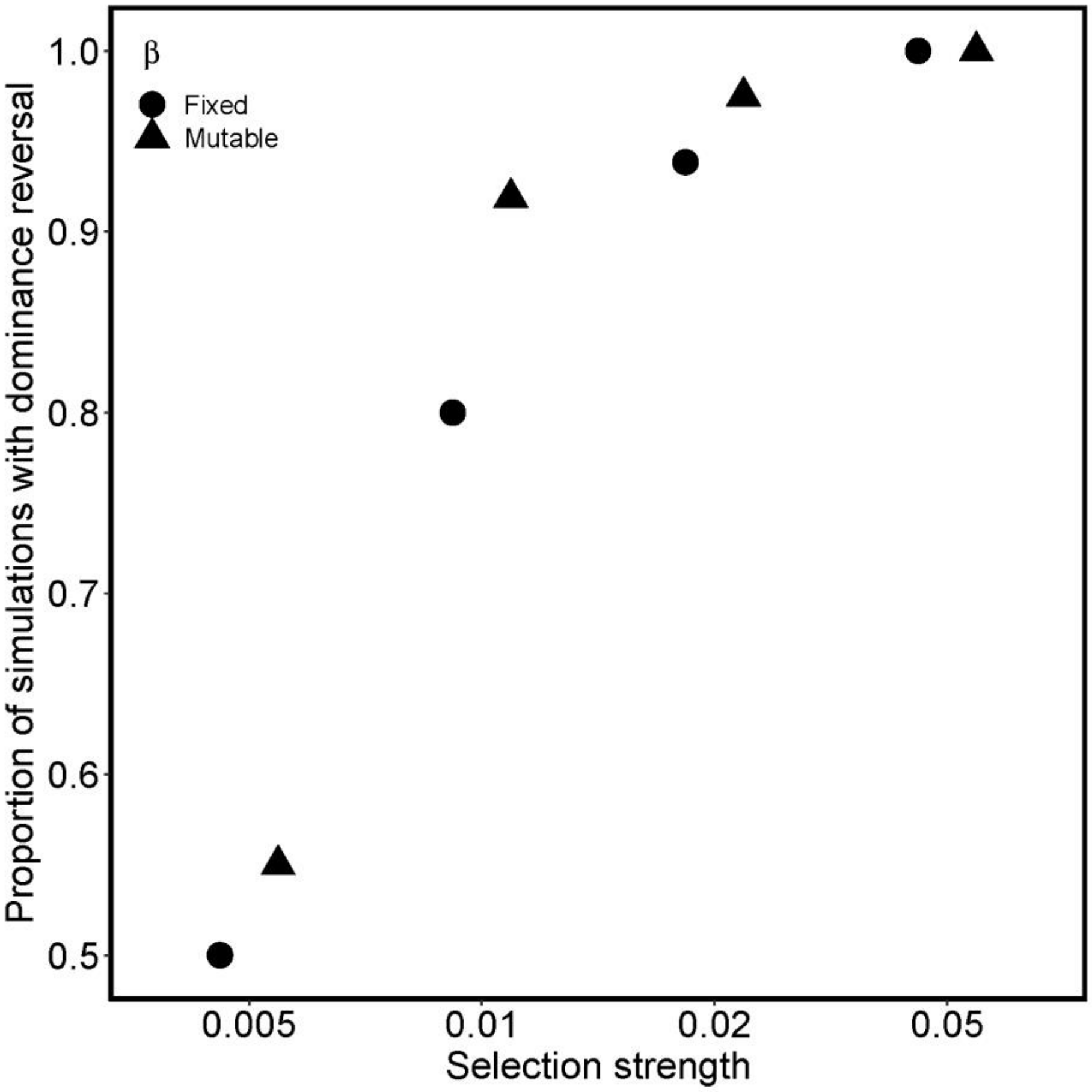

**Figure S2.3. Beneficial reversals of dominance.** Provided selection is sufficiently stronger than the mutation rate, beneficial reversals of dominance are an inevitable end state – perhaps even more likely when mutation/selection is allowed at $\beta_i$ for some strengths of selection, suggesting that epistatic interactions may be partly characteristic of sex-specific dominance modifiers. Circles and triangles represent simulations in which $\beta_i$ was fixed and mutable, respectively.

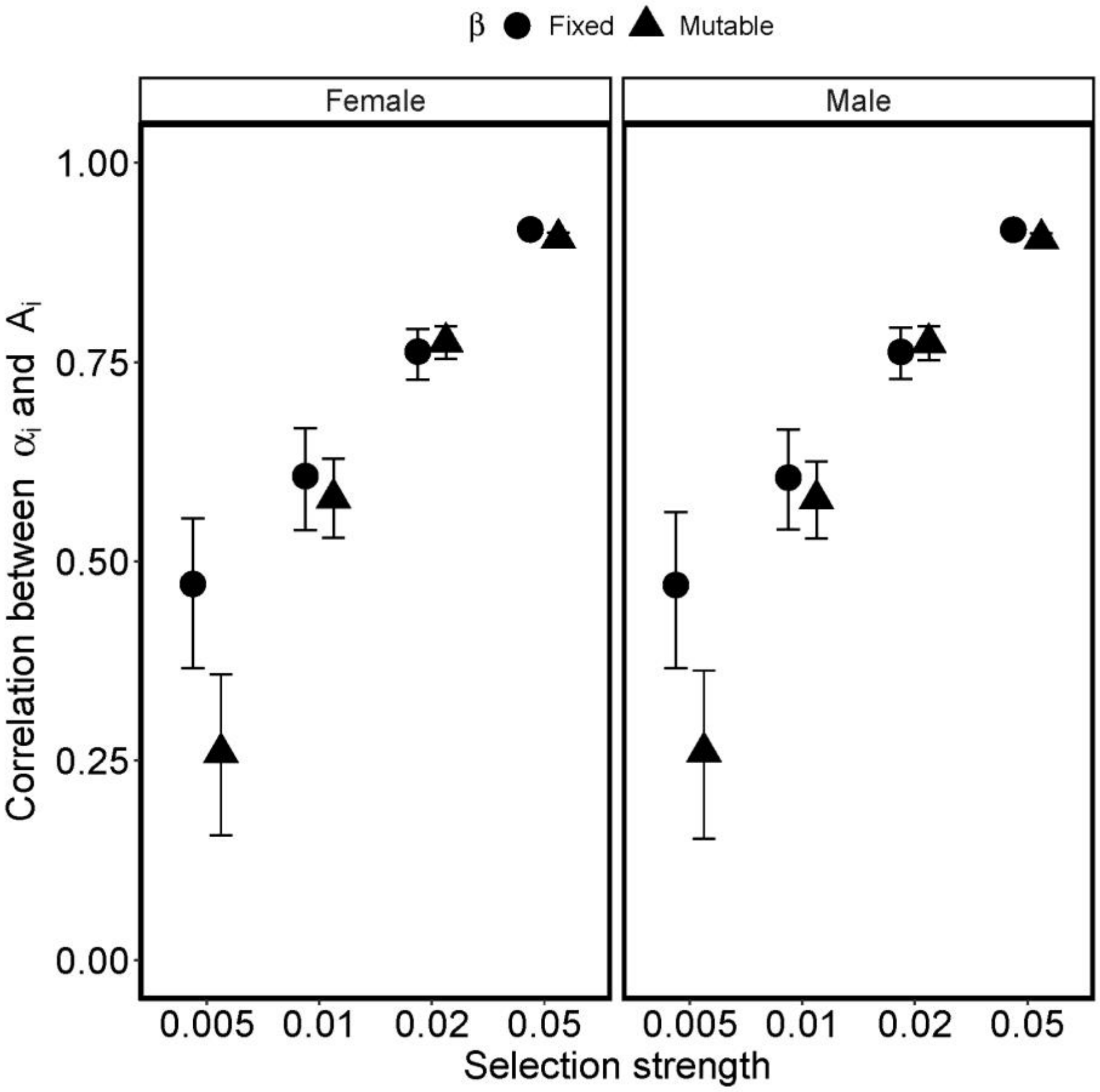

**Figure S2.4: Adaptive haplotypes.** Provided selection is sufficiently stronger than the mutation rate, selection acts to increase the frequency of adaptive haplotypes of Gene A, $\alpha_i A_i$ in both females (left) and male (right). For very weak strengths of selection ($s_f = s_m = 0.005$), this adaptive haplotype can arise more readily when $\beta_i$ is fixed. Each point represents the mean correlation (± bootstrap 95% CI) between allelic values of $A_i$ and $\alpha_i$, averaged across simulations that maintained polymorphism at $A_i$. Circles and triangles represent simulations in which $\beta_i$ was fixed and mutable, respectively.

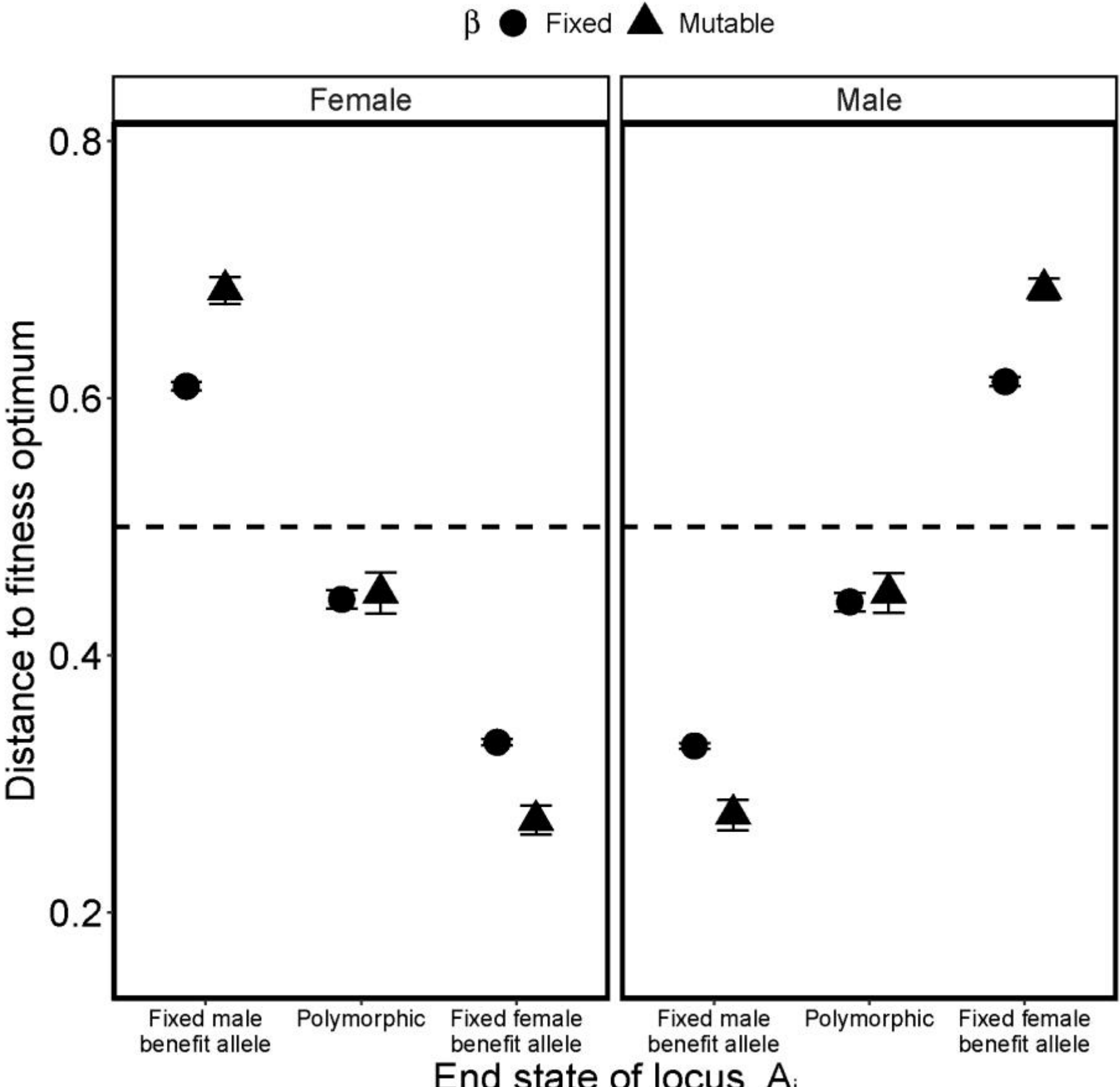

**Figure S2.5: Resolution of sexual conflict.** Mean distance to fitness optima (± bootstrap 95% CI) for simulations where $A_i$ was polymorphic, fixed for the male-benefit allele ($A_i = 1$), or fixed for the female-benefit allele ($A_i = 2$). Distance to optima were averaged across simulations with different selection coefficients ($s_f = s_m = \{0.005, 0.01, 0.02, 0.05\}$). Females (left) and males (right) were less than halfway (< 0.5; dashed line) from their fitness optima, indicating partially resolved sexual conflict relative to a codominant polymorphism. Circles and triangles represent simulations in which $\beta_i$ was fixed and mutable, respectively.

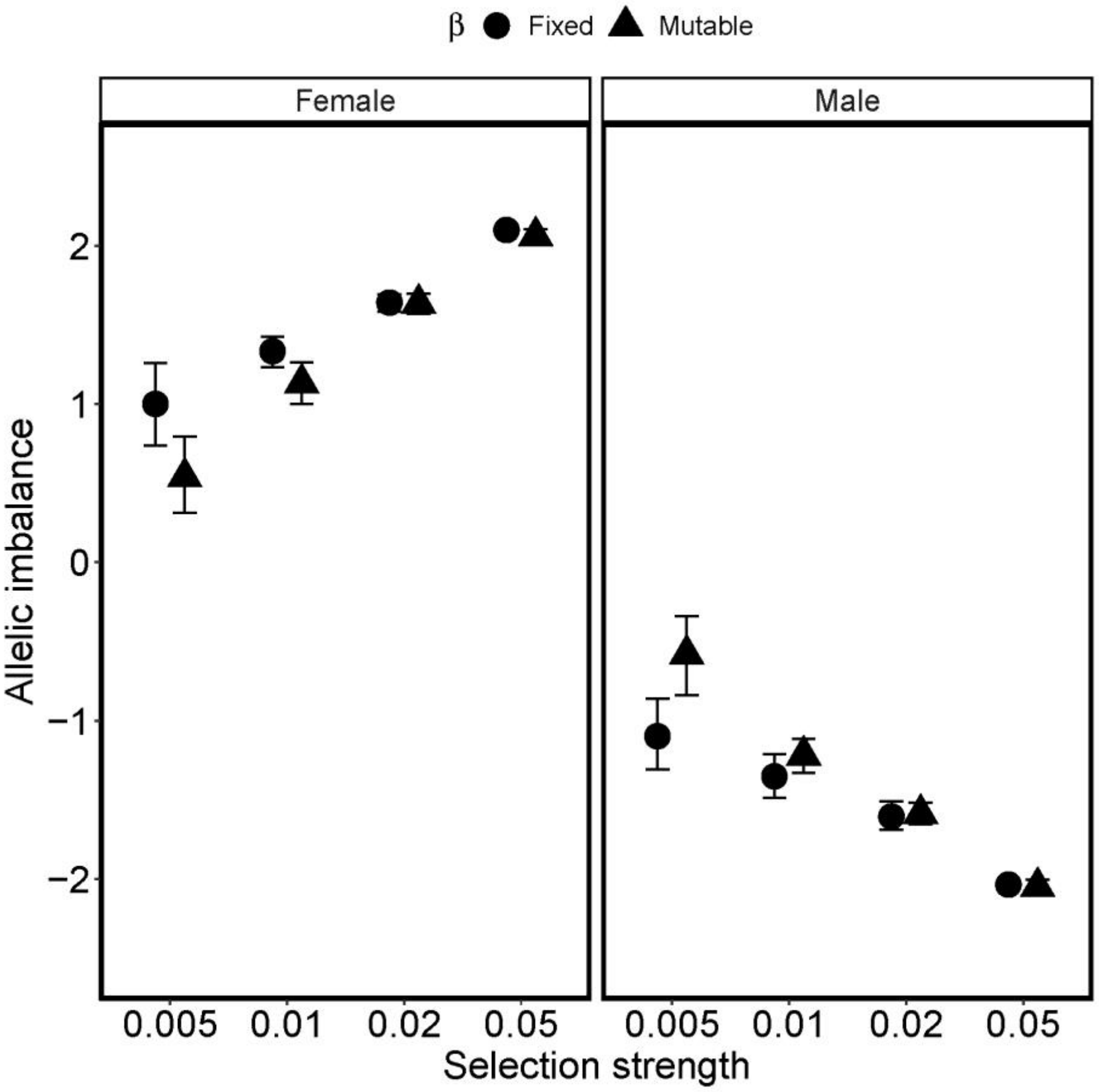

**Figure S2.6. Reversed pattern of allelic imbalance points to sex-specific dominance modifier.** Each point represents the mean (± bootstrap 95% CI) $\log_2$ fold difference in expression between $A_i = 2$ and $A_j = 1$ alleles measured in female (left) and male (right) heterozygotes, average across simulations that maintained polymorphism at $A_i$. The magnitude of the reversed allelic imbalance between the sexes increases with selection. Mutation at $\beta_i$ may hinder detection by an analysis of allele-specific expression under particularly weak strengths of selection ($s_f = s_m = 0.005$). Circles and triangles represent simulations in which $\beta_i$ was fixed and mutable, respectively.

The effect of enabling some fitness variance to stem from an epistatic interaction between Genes A and B (approximated by enabling mutation/selection at $\beta_i$) was to increase the likelihood of maintaining the focal polymorphism ($A_i$) under $s_f = s_m = \{0.01, 0.02\}$ (Figure S2.2), increase the likelihood of a beneficial reversal of dominance for fitness at $A_i$ under $s_f = s_m = \{0.005, 0.01, 0.02\}$ (Figure S2.3), reduce the ascent of beneficial haplotypes $\alpha_i A_i$ under $s_f = s_m = \{0.005\}$ (Figure S2.4), and reduce the magnitude of reversed allelic imbalance between the sexes under $s_f = s_m = \{0.005\}$ (Figure S2.6).

Although we mostly focused on the polymorphism at $A_i$, we also observe qualitative differences in the distribution of the binding site alleles ($\alpha_i$ and $\beta_i$) between simulations that maintained polymorphism at $A_i$ and those that fixed for the male-benefit or the female-benefit allele of $A_i$ (Figure S2.7, S1.8). We observe a bimodal distribution of $\alpha_i$ when polymorphism was maintained at $A_i$ (Figure S2.7). In contrast, $\alpha_i$ was unimodal when $A_i$ ended up being fixed for $A_i = 1$ or $A_i = 2$. The distribution of $\alpha_i$ skewed towards smaller values when $A_i$ was fixed for the male-benefit allele and towards larger values when $A_i$ was fixed for the female-benefit allele. The pattern observed for the distributions of $\beta_i$ was similar but not as pronounced as that for $\alpha_i$ (Figure S2.8). These results suggests that our gene-regulatory network under sexually-antagonistic selection not only maintains variation at the coding region $A_i$, but also for the binding sites $\alpha_i$ and $\beta_i$.

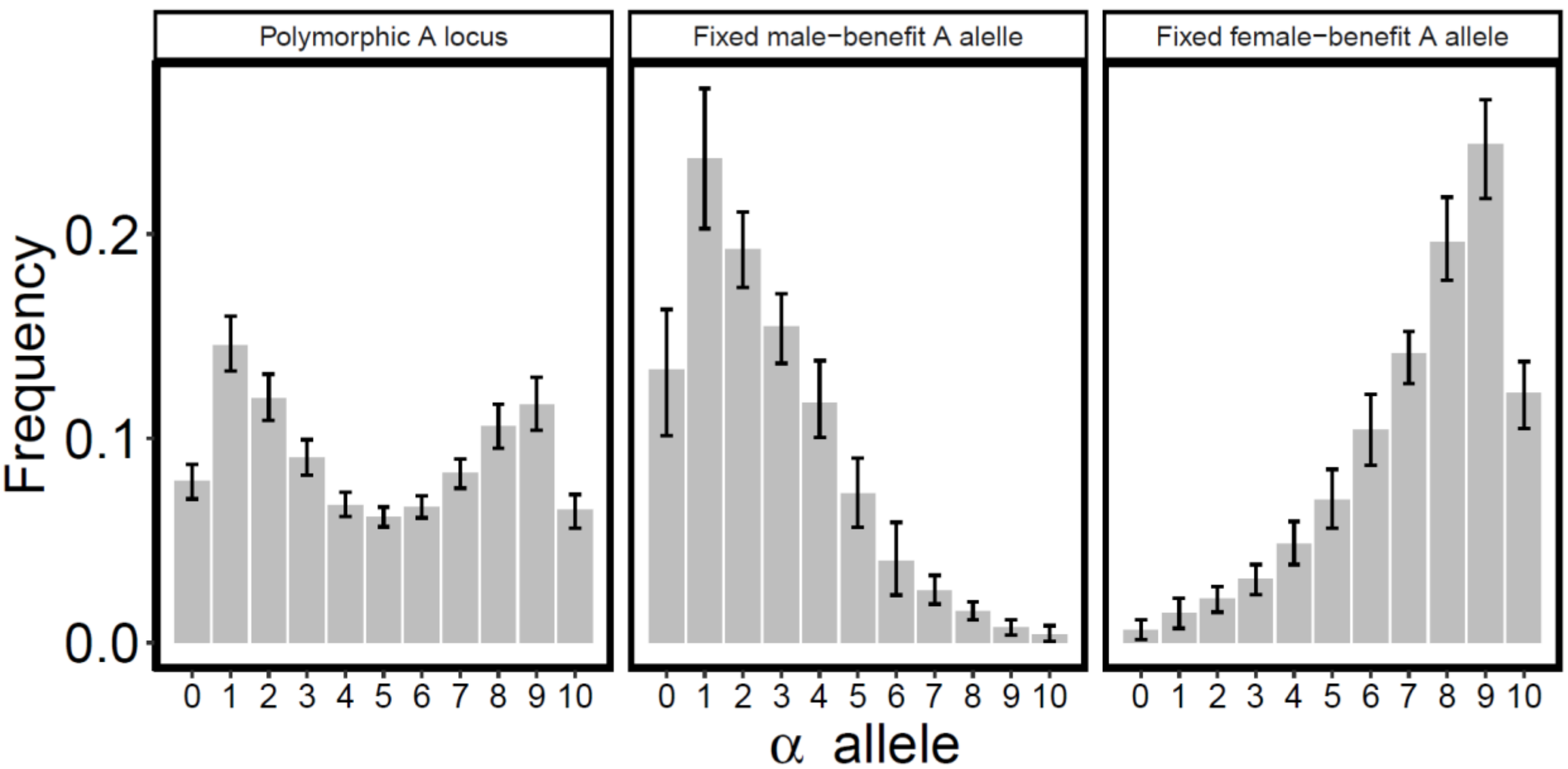

**Figure S2.7. Distribution of $\alpha_i$ alleles.** Simulations where $A_i$ was polymorphic (left), fixed for the male-benefit allele ($A_i = 1$) (middle), or fixed for the female-benefit allele ($A_i = 2$) (right) for simulations where $\beta$ was mutable. Only simulations under $s_f = s_m = \{0.02\}$ are shown. Each bar represents the mean frequency (± bootstrap 95% CI) of $\alpha_i$ alleles, averaged across simulations.

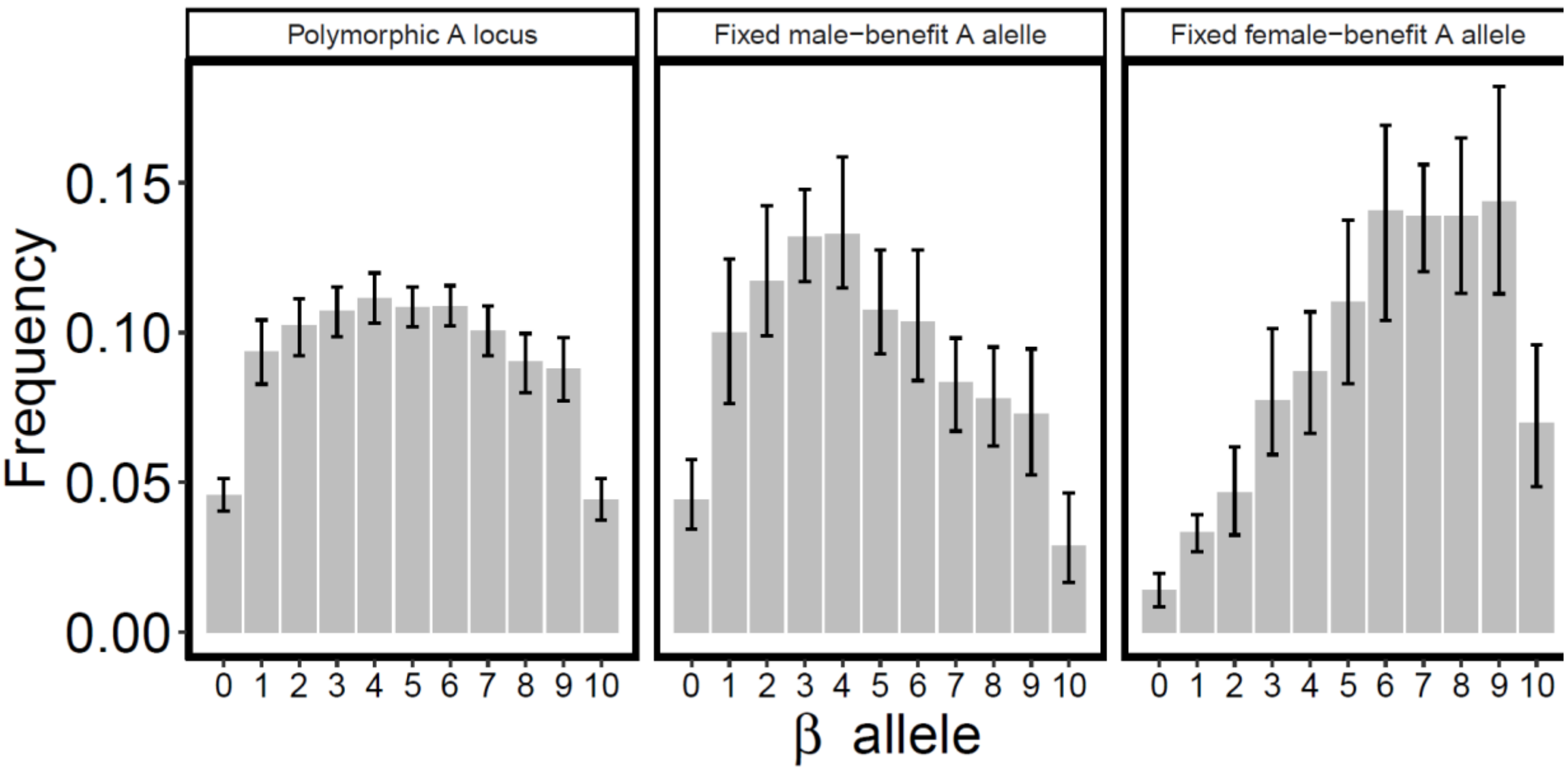

**Figure S2.8. Distribution of $\boldsymbol{\beta_i}$ alleles.** Simulations where $A_i$ was polymorphic (left), fixed for the male-benefit allele ($A_i = 1$) (middle), or fixed for the female-benefit allele ($A_i = 2$) (right) for simulations where $\beta$ was mutable. Only simulations under $s_f = s_m = \{0.02\}$ are shown. Each bar represents the mean frequency (± bootstrap 95% CI) of $\beta_i$ alleles, averaged across simulations.

## S3. Detecting dominance reversal in quantitative genetic data via dominance ordination

*Dominance ordination*

Grieshop and Arnqvist[8] developed a quantitative genetic approach capable of revealing a polygenic signature of dominance reversal by subtly modifying a rarely used method first derived by Hayman[9]. Their new method was necessary because classical and advanced quantitative genetic partitioning of fitness variance, including estimating sex-/strain-specific dominance deviations, are insufficient at definitively revealing a signature of dominance reversal, largely due to the uncertainty in connecting population/molecular genetic phenomena to quantitative genetic metrics[10]. Researchers conducting diallel crosses with the intention of assessing dominance reversals should decide prior to designing their experiment (1) whether dominance ordination is necessary for their question, (2) whether a diallel cross of sufficient size is feasible (see below).

A full diallel cross is a factorial cross among a panel of inbred strains[11]. We will present an example in which the focal context is sex, hence, each cross and reciprocal cross produces $F_1$ offspring in both contexts: male (sons) and female (daughters). The full diallel matrix includes a column and row for each of the inbred parental strains 1 through $n$, such that there are $n^2$ cells in the matrix. Some steps below require the full data set (sons and daughters) to be considered at once, and other steps require splitting the data into their context-specific matrices (i.e., separating the 'sons' and 'daughters' data). Along the diagonal of the diallel matrix are the $n$ $F_1$ inbred parental selfs (i.e., a parent strain crossed with itself), and on the off-diagonal are the $(n^2 - n)$ $F_1$ outbred crosses. Those $n^2$ cells of the full diallel matrix consist of replicated observations of trait or fitness values in $F_1$ offspring from each context (in this case, sons and daughters). Hayman[9] and Lynch and Walsh[11] emphasize that the original interpretations of this method (given below) assume no environmental and epistatic variance – importantly, the term "environmental" variance here is a quantitative genetic term that is not to be confused with "context" (i.e., niche, sex, environment, trait, life-stage, etc.), but rather refers to *unexplained/residual* variance. Grieshop and Arnqvist[8] approached this by removing that unwanted noise in the data via a reduced variance-partitioning linear model such that the non-variance-standardized residual observations still consist of additive and dominance variance, but lack environmental and epistatic variance. Note that "environmental" variance pertaining specifically to the focal context should not be removed, for example, Grieshop and Arnqvist[8] did not remove the fixed effect of 'sex' or any sex-specific effects. Note also that neither Hayman[9] nor Lynch and Walsh[11] necessarily suggest statistically removing these unwanted sources of noise, but rather state their absence as an assumption of the method; hence, Grieshop and Arnqvist's[8] solution to this is by no means the only one and may not suite other data sets. Still, it is not advisable to directly use raw data for the equations below, as the assumed absence of environmental and epistatic variance are almost certainly violated in any raw data.

The rest of the procedure is based on those residual observations that lack the unwanted noise (see above). Let $\bar{z}_{1,1}, \bar{z}_{2,2}, \bar{z}_{3,3}, \dots \bar{z}_{n,n}$ be the vector of means of the $F_1$ parental selfs, with subscripts referring to strains 1 through $n$ (Figure S3.1). For all off-diagonal cells of the diallel matrix the subscripted dam is followed by the subscripted sire, for example, a cross between a strain-1 dam and a strain-2 sire would be: $\bar{z}_{1,2}$; and the reciprocal cross: $\bar{z}_{2,1}$ (Figure S3.1). Note that the parental selfs are homozygous and that each unique cross is a replicable heterozygote. Also note that reciprocal full-siblings (e.g., $\bar{z}_{1,2}$ versus $\bar{z}_{2,1}$; Figure S3.1) are autosomally identical but have inherited their sex-chromosomes and cytoplasm from opposite strains. Below we will calculate the "array covariance" (defined below) – a value that describes each strain's relative degree of dominance in their genetic effects over that of the other strains in the diallel. The array covariances of each strain are calculated for each context (sex) separately, in this case, using separate son and daughter partitions of the diallel data set. The example shown below is for the sons of strain-1. Each strain (e.g., strain-1) has a dam- and sire-specific vector of outcrossed means (equations 2.1 and 2.2 below, respectively; Figure S3.1), and the dam- and sire-based array covariances are averaged to attain the array covariance for strain-1. For example, let $r_{\mathrm{dam}_{M^1}}$ be the vector of outbred $F_1$ male ($M$) means ($\bar{z}$) for the $n-1$ cells in which strain-1 is the dam:

$$r_{\mathrm{dam}_{M^1}}: \left[\bar{z}_{1,2_M}, \bar{z}_{1,3_M}, \bar{z}_{1,4_M}, \ldots \bar{z}_{1,n_M}\right], \tag{2.1}$$

$r_{\mathrm{sire}_{M^1}}$ be the vector of outbred $F_1$ male means for all $n-1$ cells in which strain-1 is the sire:

$$r_{\mathrm{sire}_{M^1}}: \left[\bar{z}_{2,1_M}, \bar{z}_{3,1_M}, \bar{z}_{4,1_M}, \ldots \bar{z}_{n,1_M}\right], \tag{2.2}$$

and $P_{M^1}$ be the inbred $F_1$ male means of the $n-1$ cells along the diagonal that correspond to the nonrecurrent parental selfs of the strains that strain-1 was crossed with in order to form those $r_{\mathrm{dam}_{M^1}}$ and $r_{\mathrm{sire}_{M^1}}$ crosses:

$$P_{M^1}: \left[\bar{z}_{2,2_M}, \bar{z}_{3,3_M}, \bar{z}_{4,4_M}, \ldots \bar{z}_{n,n_M}\right], \tag{2.3}$$

as indicated in Figure S3.1. The average of the two covariances $COV(r_{\mathrm{dam}_{M^1}}, P_{M^1})$ and $COV(r_{\mathrm{sire}_{M^1}}, P_{M^1})$ yields the male-specific array covariance for strain-1: $COV_{M^1}(r, P)$. Note that averaging $COV(r_{\mathrm{dam}_{M^1}}, P_{M^1})$ and $COV(r_{\mathrm{sire}_{M^1}}, P_{M^1})$ accounts for any parental imprinting that would manifest as differences between reciprocal full-siblings[11] (described above. Using the $F_1$ female ($F$) means (i.e., the daughter data) in place of the male means above – which must be measured independently – would therefore provide the female-specific array covariance for strain-1: $COV_{F^1}(r, P)$.

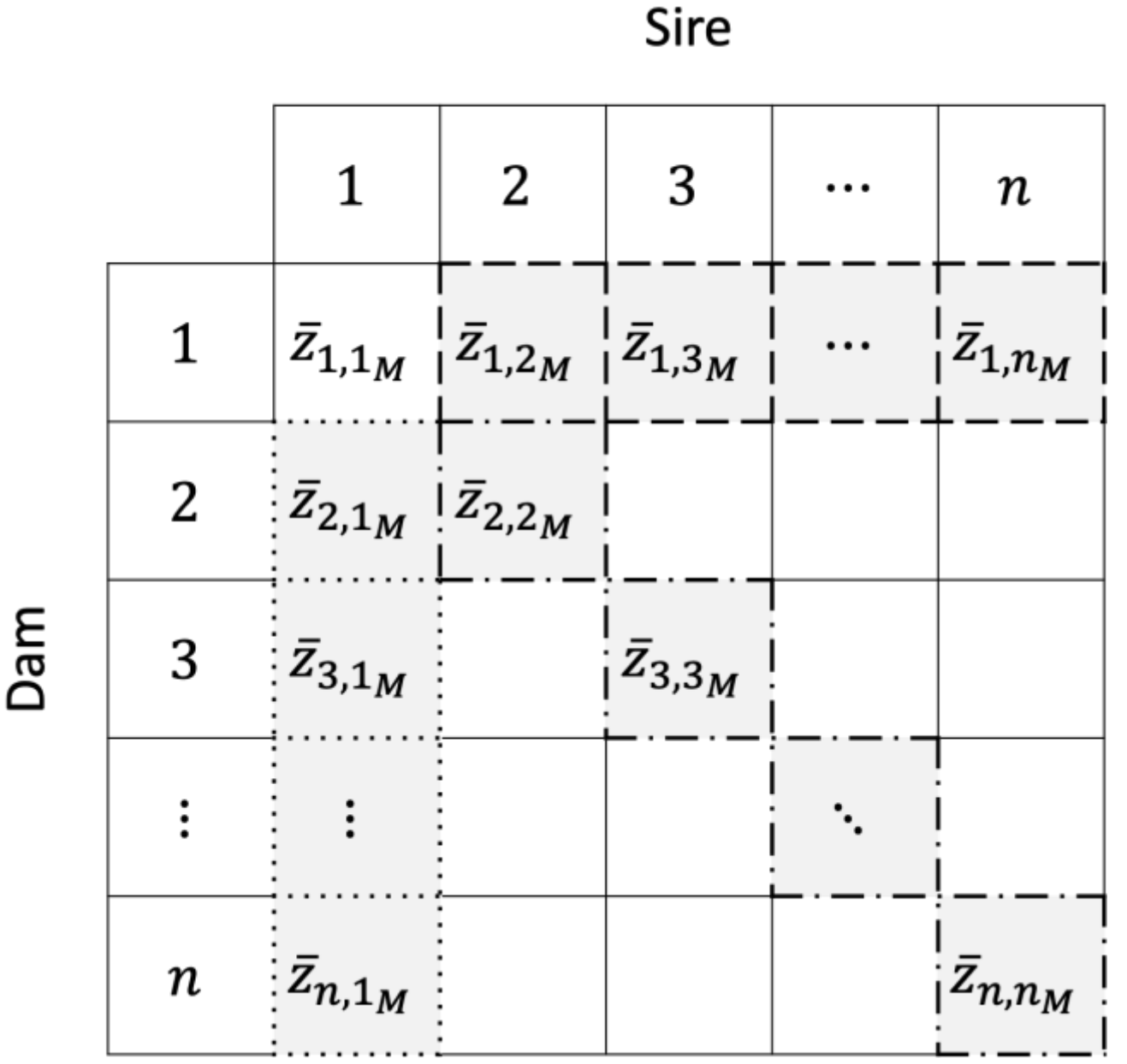

**Figure S3.1: Diallel cross schematic to aid interpretation of equations (2.1-2.5).** Sires 1 through $n$ (columns) are crossed with dams 1 through $n$ (rows). After statistically removing unwanted sources of noise (see above), replicate observations of each inbred self or outbred cross are averaged to calculated the means of each cell along the diagonal and off-diagonals, respectively. Shaded cells are those that would be used in equations (2.1-2.3), which would only be sufficient to calculate the array covariance for strain-1 males, $COV_{M^1}(r,P)$. Cells with dashed borders correspond to the vector of means in equation (2.1), cells with dotted borders correspond to the vector of means in equation (2.2), and cells with combined dashed/dotted borders correspond to the vector of means in equation (2.3). The covariance between the dashed-border and combined-border cells would be $COV(r_{\mathrm{dam}_{M^1}}, P_{M^1})$, the covariance between the dotted-border and combined-border cells would be $COV(r_{\mathrm{sire}_{M^1}}, P_{M^1})$, and their average would be the array covariance for strain-1 males, $COV_{M^1}(r,P)$. That same procedure carried out on the independently observed female data ($F$, not shown) would be used to calculate the array covariance for strain-1 females, $COV_{F^1}(r,P)$. These procedures would be conducted on the relevant cells for calculating the array covariance for the rest of the strains (2 through $n$) for males and females separately, so as to produce the vector of male and female array covariances for each strain, $W_M(r,P)$ and $W_M(r,P)$, respectively (see equations (2.4) and (2.5)).

The array covariance of each strain relative to one another provides a parametric ordering of strains in terms of whose genetic variation is dominant over whose[8] (Figure S3.2). A relatively high array covariance for a given strain indicates that its genetic variation tends to be recessive to that of the strains it was crossed with in the diallel (Figure S3.2A), as its outbred $F_1$ values are a function of who it was crossed to. Conversely, a relatively low array covariance for any strain indicates that its genetic variation tends to be dominant to the strains that it was crossed to (Figure S3.2B), as its outbred $F_1$ values are not a function of who it was crossed to. The null expectation is that these underlying allelic effects should be either additive or unconditionally dominant/recessive (i.e., not dominance-reversed), and therefore that the male- and female-specific array covariances among all strains, respectively,

$$W_M(r,P): [COV_{M^1}(r,P),\ COV_{M^2}(r,P),\ COV_{M^3}(r,P) \ldots COV_{M^n}(r,P)] \quad (2.4)$$

and

$$W_F(r,P): [COV_{F^1}(r,P),\ COV_{F^2}(r,P),\ COV_{F^3}(r,P) \ldots COV_{F^n}(r,P)], \quad (2.5)$$

should be positively correlated (Figure S3.2C). The same null expectation holds for any set of contexts for which these estimates can be independently derived: the parametric or non-parametric genetic correlation of array covariances between contexts should be positive if the majority of the underlying allelic effects are additive and/or unconditionally dominant/recessive. Grieshop and Arnqvist[8] found that both the parametric and non-parametric correlations between $W_M(r,P)$ and $W_F(r,P)$ were significantly negative (as in Figure S3.2D), rejecting the null, and indicating that the genetic variation captured within each of their strains tended to be, on average, dominant in one sex but recessive in the other (reviewed in the main text). Ideally, array covariances, dominance ordinations, and correlations between context-specific array covariances should be resampled to account for the uncertainty in their estimation.

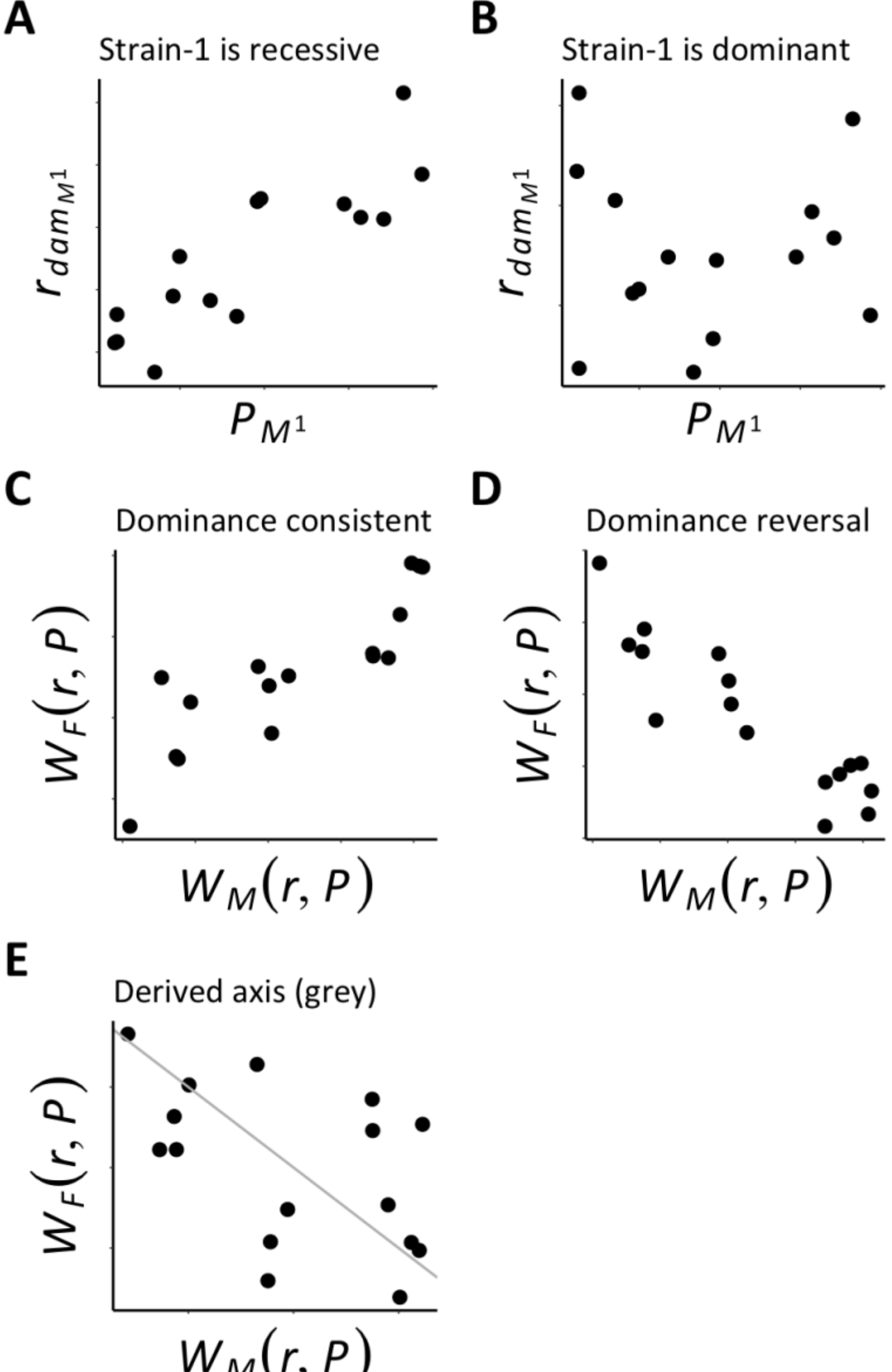

**Figure S3.2: Visual representation of some array covariance and dominance relationships.** The simulated data depict scenarios from a diallel cross among 20 strains. Panels (A) and (B) show the 19 points used to estimate the male-specific array covariance for strain-1 via dams, $COV(r_{\mathrm{dam}_{M^1}}, P_{M^1})$ (using equations 2.1 and 2.3), depicting scenarios in which the genetic variation in strain-1 is recessive and dominant, respectively, to that of other strains. When averaged with the same estimate via sires, $COV(r_{\mathrm{sire}_{M^1}}, P_{M^1})$ (using equations 2.2 and 2.3), it would provide the male-specific array covariance for strain-1, $COV_{M^1}(r, P)$. This same procedure applied to the female data would yield the female-specific array covariance for strain-1, $COV_{F^1}(r, P)$. A correlation among strains' array covariances between males, $W_M(r, P)$, and females, $W_F(r, P)$, reveals the extent to which allelic dominance tends to be consistent (C) or reversed (D) between sexes. Regardless, a derived axis (grey line, E) can ordinate strains from those harboring the greatest proportion of male-dominant alleles to those harboring the greatest proportion of female-dominant alleles.

Note that a negative correlation need not be present for this method to still provide utility in identifying a signature of dominance reversal. Consider a scenario wherein the array covariances between contexts are uncorrelated (essentially a circular cloud of points; Figure S3.2E). The cartesian coordinate system may be rotated in order to derive new dimensions[8,12]. A derived axis with a slope of -1 on the original coordinate system would represent a variable describing variation among strains from those with the most dominant alleles in one context to those with the most dominant alleles in the other context (Figure S3.2E), which could have a variety of uses. For example, using that derived axis as the response variable in a genome-wide association could potentially identify candidate regions in the genome that are associated with dominance reversal between contexts, analogous to Ruzicka et al.'s[13] derived axis of antagonistic additive genetic values. However, note that this would likely be a horribly underpowered GWAS due to the limited number of strains $n$ that can feasibly be assayed in a diallel cross (maybe ≤ 20 for most diploid organisms). For context, Ruzicka *et al.*[13] utilized 2,230 independent observations of fitness (summed across the sexes), which yielded sex-specific additive genetic breeding values for 223 unique genetic lines for their GWAS. Contrast this with Grieshop and Arnqvist's[8] similarly sized effort (3,278 independent observations of fitness, summed across the sexes) to obtain sex-specific array covariances for only 16 unique genetic lines – likely too few for a GWAS.

Lastly, regarding the interpretive framework outlined above, note that Hayman[9] derived the analytical theory to show that a regression among strains between the variance in family means, $V_r$, on the x-axis and the array covariances, $W_{r,P}$, on the y-axis should result in all strains falling along a single line with slope equal to 1, where the y-intercept indicates the population's average degree of dominance for the underlying loci, and the relative position of strains along that 1:1 line indicates their relative number of dominant/recessive alleles (see Figure 1 in Hayman[9], and Figure 20.4 in Lynch and Walsh[11]). Lynch and Walsh[11] note that measurement error (one source of environmental variance) will cause strains to deviate from this perfect 1:1 line, but that the regression coefficient should not be significantly different from one unless the inbred parental strains actually feature substantial heterozygosity or there is significant epistatic variance. However, as Grieshop and Arnqvist[8] point out, the fact that the y-intercept should indicate the average dominance coefficient of the underlying loci, and that strains should be fixed for various combinations of alleles across the genome, assumes that the dominance coefficients of alternative alleles are not conditional in any way upon the context in which they were measured, which seems unlikely, especially in light of the evidence reviewed in the main text. Indeed, Grieshop and Arnqvist's[8] inbred strains did not fall on a 1:1 line upon regressing $V_r$ and $W_{r,P}$ in either males or in females, despite having removed environmental and epistatic variance, indicating that their strains exhibited varying average degrees of dominance due to being fixed for different combinations of alleles. Thus, Grieshop and Arnqvist[8] focused on comparing the array covariances between contexts (males and females); after all, if the relative position of strains along the 1:1 line should indicate their relative number of recessive alleles then so should their relative position along either the x- or y-axis alone, the latter of which carries the intuitive logic outlined above and in Figure S3.2. That said, it is possible that strains' relative values along the 1:1 line between $V_r$ and $W_{r,P}$, rather than along the latter axis alone, prove useful for some data sets.

## REFERENCES FOR SUPPORTING INFORMATION

1. Porter, A. H., Johnson, N. A. & Tulchinsky, A. Y. A New Mechanism for Mendelian Dominance in Regulatory Genetic Pathways: Competitive Binding by Transcription Factors. *Genetics* **205**, 101–112 (2017).

2. Rosen, G. A. *et al.* Dynamics of RNA polymerase II and elongation factor Spt4/5 recruitment during activator-dependent transcription. *Proc. Natl. Acad. Sci.* **117**, 32348–32357 (2020).

3. Hochberg, G. K. A. *et al.* A hydrophobic ratchet entrenches molecular complexes. *Nature* **588**, 503–508 (2020).

4. Stewart, A. J., Hannenhalli, S. & Plotkin, J. B. Why transcription factor binding sites are ten nucleotides long. *Genetics* **192**, 973–985 (2012).

5. Verhulst, E. C., van de Zande, L. & Beukeboom, L. W. Insect sex determination: it all evolves around transformer. *Curr. Opin. Genet. Dev.* **20**, 376–383 (2010).

6. Chen, J., Nolte, V. & Schlötterer, C. Temperature Stress Mediates Decanalization and Dominance of Gene Expression in Drosophila melanogaster. *PLOS Genet.* **11**, e1004883 (2015).

7. Mishra, P., Barrera, T. S., Grieshop, K. H. & Agrawal, A. F. Cis-regulatory variation in relation to sex and sexual dimorphism in Drosophila melanogaster. *bioRxiv* (2022).

8. Grieshop, K. & Arnqvist, G. Sex-specific dominance reversal of genetic variation for fitness. *PLOS Biol.* **16**, e2006810 (2018).

9. Hayman, B. I. The Theory and Analysis of Diallel Crosses. *Genetics* **39**, 789–809 (1954).

10. Huang, W. & Mackay, T. F. C. The Genetic Architecture of Quantitative Traits Cannot Be Inferred from Variance Component Analysis. *PLOS Genet.* **12**, e1006421 (2016).

11. Lynch, M. & Walsh, B. Genetics and analysis of quantitative traits. (1998).

12. Berger, D. *et al.* Intralocus Sexual Conflict and Environmental Stress. *Evolution* **68**, 2184–2196 (2014).

13. Ruzicka, F. *et al.* Genome-wide sexually antagonistic variants reveal long-standing constraints on sexual dimorphism in fruit flies. *PLOS Biol.* **17**, e3000244 (2019).